\documentclass[11pt,onecolumn,showpacs,superscriptaddress,preprintnumbers,floatfix,nofootinbib]{revtex4}
\usepackage{epsfig}
\usepackage{amsfonts}
\usepackage{mathrsfs}
\usepackage{latexsym}
\usepackage{color}
\usepackage{amsmath,amssymb}
\usepackage{verbatim}
\definecolor{red}{rgb}{1,0,0}

\def\dv{v}
\def\dk{\kappa}

\def\bea{\begin{eqnarray}}
\def\eea{\end{eqnarray}}
\def\be{\begin{equation}}
\def\ee{\end{equation}}
\def\ba{\begin{array}}
\def\ea{\end{array}}
\def\nn{\nonumber}
\def\sfa{single--field approximation}

\begin{document}

\title{Functional renormalization of $N$ scalars with $O(N)$ invariance}
%\date{}
%\maketitle
\bigskip

\author{R.~Percacci} \email{percacci@sissa.it}
\affiliation{
International School for Advanced Studies,
via Bonomea 265, 34136 Trieste, Italy}
\affiliation{INFN, Sezione di Trieste, Italy}
\author{M. Safari} \email{safari@sissa.it, mahsafa@gmail.com}
\affiliation{
International School for Advanced Studies, via Bonomea 265, 34136 Trieste, Italy}

\begin{abstract}
We discuss general theories of $N$ scalar fields with $O(N)$ symmetry.
In addition to the standard case of linearly realized symmetry
there are also examples that carry nonlinear realizations,
with the topology of a cylinder $R\times S^{N-1}$ or a sphere $S^N$.
We write flow equations for the theory in the second order of the
derivative expansion in the background field
and discuss the properties of scaling solutions with vanishing potential.
\end{abstract}
\maketitle

%%%%%%%%%%%%%%%%%%%%%%
\section{General scalar theories with $O(N)$ symmetry}
%%%%%%%%%%%%%%%%%%%%%%

The most economical nontrivial realization of $O(N)$ symmetry in a scalar field theory
makes use of $N-1$ fields. It corresponds to a nonlinear sigma model
with values in the $(N-1)$--sphere $S^{N-1}=O(N)/O(N-1)$, where the radius of the sphere
is treated as a fixed parameter.
In many applications it is preferable to turn the radius into a dynamical field.
Then, one can reparameterize the theory in terms of $N$ scalar fields $\phi^a$,
$a=1,\ldots,N$, transforming linearly under $O(N)$.
Aside from the advantage of working with simpler, linear transformations,
such models are also better behaved as quantum field theories.

In this paper we will consider the renormalization of $O(N)$-invariant
models with $N$ fields but without relying on the linear structure.
There are two aspects to this.
As a first step, the metric in the target space could differ from the flat one,
but maintaining the topology $R^N$.
This happens for example if one allows interactions of the form
$Z(\phi^2)\partial_\mu\phi^a\partial^\mu\phi^a$ (in four dimensions, this requires
operators of dimension at least six).
As is always the case in mechanics, the term quadratic in time derivatives defines
a metric in the configuration space. Here one has a
conformally flat metric $Z(\phi^2)\delta_{ab}$,
so the target space becomes effectively a Riemannian manifold
and one could treat these theories as general nonlinear sigma models.
However, there still exist global coordinates such that the fields tranform in
the familiar linear way under $O(N)$, and for this reason we will still say that
the fields carry a linear realization of the global symmetry.
The second aspect is the topology.
By removing the origin, one can change the topology into
that of a cylinder $S^{N-1}\times R$,
and by adding a single point at infinity one can change the topology into that of a sphere $S^N$.
In both cases there are now {\it local} coordinate system such that the fields transform in
the usual linear way, but such coordinates cannot be extended to cover the whole space.
In fact, neither the cylinder nor the sphere are linear spaces,
so in these cases we will say that the fields carry nonlinear realizations.
Such models can be viewed as limits of deformations of linear models.

The motivation for studying these models comes at least in part from Higgs physics.
Consider a linear model in the spontaneously broken phase where
the field has a nonvanishing vacuum expectation value (VEV).
Perturbative analysis of the fluctuations around the vacuum reveals
the presence of $N-1$ massless modes, the Goldstone bosons,
and one massive mode, corresponding to the radius.
Suppose we observe such a spectrum, and furthermore suppose that the
scattering amplitudes exhibit $O(N)$ invariance.
This is still not enough to validate the simple linear scalar theory:
the target space could deviate significantly from flatness
for fields that are either much larger or much smaller than the VEV or both.

Various nonlinear models of this type have been used in phenomenology.
One class of models is based on the observation that the VEV of the Higgs
is the only parameter in the standard model that breaks scale invariance.
Instead of an explicit breaking, as in the SM, it is tempting to think that
scale invariance is spontaneously broken.
Then, the radius, parametrized as $\rho=e^\sigma$, where $-\infty<\sigma<\infty$
can be interpreted as a dilaton.
In this model the topology of the target space is a cylinder,
and the symmetry of the action is enhanced by scale invariance.
Various applications of similar ideas to the standard model have been discussed
in \cite{buchmuller}.

Another class of models that have enjoyed some popularity recently,
the so-called composite Higgs models, are based on the assumption that
the four real degrees of freedom of the Higgs doublet are the Goldstone
bosons resulting from the breaking of some global symmetry group $G$
to a subgroup $H$ \cite{georgi}. The minimal model of this type, giving rise to the Higgs
doublet as the only Goldstone bosons, while preserving custodial symmetry,
corresponds to $SO(5)$ spontaneously breaking to $SO(4)$
\cite{contino} (see also \cite{serone} and references therein).
In this case the topology of the target space would be a four-sphere.
Note that in both classes of models the radial mode
can be interpreted as a Goldstone boson,
on the same footing with the angular degrees of freedom.
In both cases one loses the simple linear realization of the symmetry,
but this may not be easily seen in perturbation theory around the VEV.

It is expected that these nonlinear models break down at some energy
and that they are to be treated as low energy effective field theories.
In this paper we shall discuss the renormalization of such theories
using functional renormalization group methods.
One powerful general method of studying quantum field theories is to integrate out
quantum fluctuations gradually, one momentum shell at the time, beginning from
some ultraviolet scale $\Lambda_{UV}$ down to some scale $k$.
The original implementation of this idea by Wilson gives rise to a $k$--dependent
Wilsonian action $S_k$ \cite{wilson}.
An exact functional equation was derived in \cite{wh,polchinski} and applied
to the study of scalar theories in \cite{hasenfratz}, and later also to general $O(N)$ models \cite{liao_polonyi}.
For many purposes it has proven more convenient to study instead
the $k$--dependence of a one--particle irreducible functional $\Gamma_k$,
called the ``Effective Average Action'' (EAA),
which is defined exactly like the effective action, but with a smooth cutoff
in the functional integral (see \cite{reviews,btw,delamotte} for introductory reviews).
More precisely, in the functional integral one adds (by hand) to the action
a term quadratic in the fields that in momentum space looks like
$\Delta S_k=\int dp\, \phi^a(-p) R_k(p^2) \phi^a(p)$,
where $R_k(p^2)$ is a monotonically decreasing function with $R_k(0)=k^2$,
$R_0(p^2)=0$ and tending rapidly to zero for $p^2>k^2$.
The role of this term is to suppress the contribution to the functional integral of
the modes with $p^2<k^2$, so it is referred to as the infrared cutoff.
The nice property of this functional is that it satisfies a simple functional equation
\cite{wetterich,morris1}
\be
\label{ERGE}
k\frac{\partial\Gamma_k}{\partial k}=
\frac{1}{2}{\rm Tr}\left(\frac{\delta^2\Gamma_k}{\delta\phi^2}+R_k\right)^{-1}
k\frac{\partial R_k}{\partial k}\ .
\ee
This functional equation specifies the $k$--dependence of the EAA,
so the right-hand-side can be considered as a ``beta functional'' of the theory.
It contains the beta functions of all the couplings that are present in the EAA.
Due to the fall--off properties of $R_k$, the trace on the r.h.s. is finite.
One can therefore take the equation as a basis for defining a quantum field theory:
given the field content and the symmetries, one can calculate the r.h.s. of the equation.
It defines a flow on the ``theory space'' parameterized by the functionals $\Gamma_k$.
Choosing some initial condition, one can then, at least in principle,
calculate the effective action by solving the flow in the limit $k\to0$.

It is of course impossible to do this in practice.
A useful approximation method of this equation is the derivative expansion,
where one retains terms up to some fixed number of derivatives.
This is well-motivated in statistical mechanics in
applications to the theory of phase transitions
and in particle physics for the study of low energy effective field theories.
For a simple scalar theory, the lowest order of this expansion is the ``Local Potential
Approximation'' (LPA) where one retains a fixed quadratic kinetic term
and the only interactions are given by a potential $V(\phi^2)$.
Inserting this ansatz in \eqref{ERGE} one obtains a differential equation for
the flow of the potential, that can be studied to analyze the scaling solutions
\cite{litim1,cododo}.
A slightly improved version of the LPA, sometimes called LPA', takes into account the running
of a wave function renormalization constant $Z$.
This yields improved values for the critical exponents and also gives the anomalous dimension
\cite{tl,canet1}.

In the second order of the derivative expansion one retains the most general terms
that are $O(N)$--invariant and contain at most two derivatives:
\be
\label{linearaction}
\Gamma_k(\phi)=\int d^dx\left[
\frac{1}{2}Z(\phi^2)\partial_\mu\phi^a\partial^\mu\phi^a
+Y(\phi^2)\phi^a\partial_\mu\phi^a\phi^b\partial^\mu\phi^b
+V(\phi^2)\right]\ ,
\ee
where $Z$ and $Y$ are arbitrary, $k$--dependent functions of $\phi^2$.
(We do not indicate the $k$--dependence explicitly for typographical simplicity.)
The scaling solutions of these models in $d=3$ have been studied in \cite{morris2},
using standard methods of linear realizations.
Since the functions $Z(\phi^2)$ and $Y(\phi^2)$ define a curved
target space geometry, it makes sense to study such models also with methods that had been
previously applied to the RG of nonlinear sigma models \cite{codello1}.
Here, however, the situation is more complicated.
In typical nonlinear sigma models the symmetry group acts transitively on the
target space and therefore the only invariant potential is a constant.
In the case we want to discuss, the potential need not be a constant
(though it is constant on the orbits $S^{N-1}$) so we have to discuss its flow too.

In this paper we will lay the foundations for a study of these theories in the
second order of the derivative expansion
(the case $N=1$ has already been discussed to next order of the derivative expansion
\cite{canet2}).
For the quantization of the nonlinear sigma models we shall use the background field method.
After choosing a background field $\bar\varphi^i$
any other field that is not too distant from it
can be parametrized in terms of normal coordinates $\bar\xi^i(x)$.
The action can then be expanded in powers of $\bar\xi$
and the cutoff is quadratic in $\bar\xi$, as we shall describe later.
This method has the advantage of preserving invariance under
simultaneous coordinate transformations of $\bar\varphi$ and $\bar\xi$,
but it has the drawback that the EAA becomes a functional
of two field $\Gamma(\bar\varphi,\bar\xi)$.
In principle we would therefore have to allow many more invariants than in a
functional of $\phi$ alone.
In this paper we will not consider such complications and study the flow of
background functionals only, which are the most important ones in the analysis
of the phase structure of the theory.
Regarding the wave function renormalization of $\bar\xi$, we shall make
two separate approximation: one is to assume that it is $k$-independent
({\it i.e.} that the anomalous dimension of $\bar\xi$ is zero),
the other is to assume that the wave function renormalization of
$\xi$ can be approximated by the wave function renormalization of the background.
We will see that even though we go to second order in derivatives of the background field,
in both cases the results are close to those of the LPA in the linear formalism.

After deriving the flow equations for the theory, we will establish the existence
of scaling solutions with three different target space topologies:
the linear topology $R^N$, the cylinder $S^{N-1}\times R$ and the sphere $S^N$.
We will only discuss scaling solutions with constant potential
and determine in each case the set of relevant deformations.
At least in the linear case and in dimensions $2<d<4$ there are also
scaling solutions with nontrivial potential.
We will not repeat here the analysis of these solutions, but it will be useful
at some later stage to check that the geometrical methods used here agree with the
standard ones used in \cite{morris2}.

%%%%%%%%%%%%%%%%%%%%%%%%%%%%%%%%%%%%%%%%%%%%%
\section{Flow equations}
%%%%%%%%%%%%%%%%%%%%%%%%%%%%%%%%%%%%%%%%%%%%%

We will study the RG flow of theories with action of the form \eqref{linearaction},
but we will not assume that the topology of the target space is $R^N$.
In other words, the action \eqref{linearaction} may only be valid locally in the target space.
For this reason we shall use a formalism that is manifestly invariant under
coordinate transformations in the target space.
This formalism has been developed in the context of nonlinear sigma models
\cite{honerkamp} and has been applied previously to the RG flow of
these theories \cite{codello1}.
Although one could actually work with completely arbitrary coordinates,
it is convenient to assume that the coordinate system is adapted
to the action of the group $O(N)$, in the sense that one coordinate $\rho$
parametrizes the different orbits of the group and the remaining
$N-1$ coordinates $\chi^\alpha$ are coordinates within the orbits $S^{N-1}$.
Let us consider first what happens in the linear case.
The coordinate transformation from the adapted coordinates to the linear
coordinates $\phi^a$ is of the form $\phi^a=\rho\,\hat\phi^a(\chi^\alpha)$,
with $\rho=\sqrt{\sum_i\phi_i^2}$.
Then the action \eqref{linearaction} may be rewritten in the form
\be
\label{L_Simple}
\Gamma_k=\int d^dx\left[\frac{1}{2}\,J(\rho)\,\partial_{\mu}\rho\partial^{\mu}\rho
+\frac{1}{2}\,K(\rho)\,g_{\alpha\beta}\partial_{\mu}\chi^{\alpha}\partial^{\mu}\chi^{\beta}
+V(\rho)\right]\ ,
\ee
where $g_{\alpha\beta}$ is the metric on the unit $(N-1)$--dimensional sphere,
$J=Z+2\rho^2Y$, $K=\rho^2Z$.
In the following we will assume that the EAA has the form \eqref{L_Simple},
without assuming that it has been derived from \eqref{linearaction}.
Thus the coordinate $\rho$ need not have the meaning of ``radius''.

The terms in \eqref{L_Simple} with two derivatives can be rewritten as
\be
\frac{1}{2}G_{ij}\partial_{\mu}\varphi^{i}\partial^{\mu}\varphi^{j}
\ee
where $\varphi^i = (h, \chi^{\alpha})$ are adapted coordinates on the target space and
\be
\label{metric}
G_{ij}= \left(
\ba{c|c}
J(\rho) & \\ \hline &K(\rho)\,g_{\alpha\beta}
\ea
\right)
\ee
is the metric. Note that this is a Euclidean Robertson--Walker metric
with spherical ``spatial'' sections.
The coordinate transformation
\be
\label{coordtransf}
h(\rho)=\int_0^\rho \sqrt{J(x)}dx
\ee
brings the metric to standard form with $J=1$.
Notice that the field $h$ is canonically normalized and therefore has dimension $\frac{d-2}{2}$.
We will use this field reparametrization later on, but first we have to derive
the beta functions of $J$, $K$ and $V$.

It was found in \cite{codello1} that in the nonlinear sigma models
the flow of the metric $G_{ij}$ is governed by the Ricci tensor of $G_{ij}$.
Here we have to take into account the additional complications due to the presence
of a nontrivial potential.
We shall use the background field method.
After choosing a background field $\bar\varphi^i(x)$,
any other field $\varphi^i(x)$ that is not too distant from $\bar\varphi^i(x)$
can be parametrized in terms of the normal coordinates $\xi^i(x)$ as
$\varphi^i(x)=Exp_{\bar\varphi(x)}\xi^i(x)$.
Quantization produces an EAA that is a functional of two fields $\Gamma_k(\bar\varphi;\xi)$.
We can expand the EAA in powers of $\xi$:
\be
\Gamma_k(\bar\varphi;\xi)=\bar\Gamma_k(\bar\varphi)
+\Gamma_k^{(1)}(\bar\varphi,\xi)
+\Gamma_k^{(2)}(\bar\varphi,\xi)+\ldots
\ee
where $\Gamma_k^{(n)}(\bar\varphi;\xi)$ contains $n$ powers of $\xi$.
In particular $\bar\Gamma_k(\bar\varphi)\equiv\Gamma_k^{(0)}(\bar\varphi)=\Gamma_k(\bar\varphi;0)$ depends only on the background.
Throughout this paper we will work in a ``single field truncation'', which means
that the r.h.s. is assumed to be the Taylor expansion of some functional of the full field $\varphi=Exp_{\bar\varphi}\xi$ (and one can call that functional $\bar\Gamma_k(\varphi)$).
Furthermore, we assume that this functional has the form (\ref{L_Simple}).
By inserting this ansatz in the flow equation \eqref{ERGE} we will derive
``beta functionals'' for $J$, $K$ and $V$, which we will call
\be
\label{etas}
\zeta_J=\frac{d}{dt}\log J\ ,\qquad
\zeta_K=\frac{d}{dt}\log K\ ,\qquad
\zeta_V=\frac{d}{dt}\log V\ ,
\ee
where $t=\log k$. To evaluate the r.h.s. of \eqref{ERGE} we start by Taylor expanding (\ref{L_Simple})
to second order in $\xi$.
From the two--derivative terms we get
\be
\frac{1}{2}\int \!\! d^{d}x \;\xi^i \big(- G_{ij}\nabla^2 -M_{ij}\big) \xi^j\ ,
\ee
where $\nabla_\mu\xi^i=\partial_\mu\xi^i+\partial_\mu\varphi^k\Gamma_k{}^i{}_j\xi^j$ and
$M_{ij} = \partial_{\mu}\varphi^m \partial^{\mu}\varphi^n R_{imjn}$.
Here $\Gamma_k{}^i{}_j$ are the Christoffel symbols of the metric $G_{ij}$ and
$R_{imjn}$ is its Riemann tensor, whose nonzero components are
\bea
R_{\alpha 0 \beta 0} &=& \frac{K K' J'+K^{\prime 2}J-2 K K'' J}{4 K^2 J} \,G_{\alpha\beta} \label{Riemann0}
\\
R_{\alpha \gamma \beta \delta} &=& \frac{4K J-K^{\prime 2}}{4 K^2 J} \;\left(G_{\alpha\beta}G_{\gamma\delta}-G_{\alpha\delta}G_{\gamma\beta}\right) \ .
\label{Riemann}
\eea
The expansion of the potential is
\be
V(h) = V(\bar{h}) + V'(\bar{h})\left(\xi^0-\frac{1}{2}\Gamma^{0}_{ij}\,\xi^i \xi^j+\cdots\right) +\frac{1}{2} V''(\bar{h})\,\left(\xi^0 -\frac{1}{2}\Gamma^{0}_{ij}\,\xi^i \xi^j+\cdots\right)^2 +\cdots
\ee
where the non zero components of $\Gamma^{0}_{ij}$ are
$\Gamma^{0}_{\alpha\beta} = -\frac{K'}{2J}\,g_{\alpha\beta}$, $\Gamma^{0}_{00} = \frac{J'}{2J}$.
This gives a contribution to $\Gamma_k^{(2)}$ of the form
$\frac{1}{2}\int dx\,\xi^i S_{ij} \xi^j$,
with
\be
S_{ij}=V'' \,\delta^{0}_{i} \delta^{0}_{j}- V' \,\Gamma^{0}_{ij}=
(V''-V' J' /2J)\,\delta^{0}_{i} \delta^{0}_{j}+(V'K'/2J) \;g_{ij}\ .
\ee
(Here, we think of $g_{ij}$ as a $N\times N$ matrix with $g_{00}=g_{0i}=0$).
Altogether the second order expansion of (\ref{L_Simple}) yields
\be
\Gamma_k^{(2)}(\bar\varphi,\xi)=\frac{1}{2}\int \!\! d^{d}x \;\xi^i \big(-G_{ij} \nabla^2 -M_{ij}+S_{ij}\big) \xi^j
\label{tony}
\ee
The cut-off function is then conveniently chosen as
\be
\label{cut-off}
({\cal R}_k)_{ij} = G_{ij} R_{k}(z) %=  I\,g_{ij}\, R_{k}(z) + J \,  \delta^{0}_{i} \delta^{0}_{j}\, R_{k}(z)
= \left(
\ba{c|c}
J R_{k}(z)& \\ \hline &K\,g_{\alpha\beta}R_{k}(z)
\ea
\right)
\ee
where $z$ stands for the covariant Laplacian $-\nabla^{2}$.
We now have all the pieces that enter in the r.h.s. of the FRGE.
Adding the cut-off, the quadratic action can be written as
\bea
\Gamma^{(2)}+\Delta S_k = \frac{1}{2}\int \!\! d^{d}x \;
\xi^i \big({\cal P}_{ij}(-\nabla^2)-M_{ij}+S_{ij}\big) \xi^j
\eea
where ${\cal P}_{ij}=G_{ij}P_k$ and $P_k(z)=z+R_k(z)$.
The derivative of the cut-off function with respect to $t$ is
\be
\label{dcut}
\dot{\cal R}_{ij}\equiv \frac{d {\cal R}_{ij}}{dt}
= \left(
\ba{c|c}
 J\left(\dot{R}_{k}+\zeta_J R_{k}+R'_{k}\dot{z}\right)&
\\
\hline &\,K g_{\alpha\beta}\left(\dot{R}_{k}+\zeta_K R_{k}+R'_{k}\dot{z}\right)
\ea\right)
\ee
where $\dot{R}_{k}(z)=\partial_t R_{k}(z)$, $R'_{k}(z)=\partial_z R_{k}(z)$.
The terms involving $\dot{z}$ will give no contribution.

In order to calculate the beta functions for $V$, $J$, $K$ we have to extract
from the r.h.s. of the FRGE the terms that contain either no derivatives
or two derivatives of the background field.
Recalling that $M$ contains two derivatives of the background field,
the r.h.s. of the FRGE can be expanded in $M$ as
\be
\label{expans}
\frac{1}{2}\mathrm{Tr}\,\frac{\dot{\cal R}_k}{{\cal P}_k-M+S}=
\frac{1}{2}\mathrm{Tr}\,
\left(1+({\cal P}_k+S)^{-1}M+\ldots\right)({\cal P}_k+S)^{-1}\dot{\cal R}_k
\ee
Only the first two terms are needed for our calculation,
which is described in the Appendix.

Note that, in $\dot{\cal R}_k$, $-\zeta_J$ and $-\zeta_K$
play the role of anomalous dimensions of the fields.
A more treatment, which goes beyond the single-field truncation,
would consist in replacing the factors of $J$ and $K$ contained in the metric $G_{ij}$
in \eqref{tony} with independent wave function renormalization constants
$Z_J$ and $Z_K$.
Then in \eqref{dcut} $\zeta_J$ and $\zeta_K$ would be replaced by
$-\eta_J=\frac{d\log Z_J}{d\log k}$ and $-\eta_K=\frac{d\log Z_K}{d\log k}$.
These ``genuine'' anomalous dimensions would then be obtained from the $t$-derivative
of the two-point functions $\langle\xi^i\xi^j\rangle$.
See \cite{fwz} for such a calculation in a nonlinear sigma model.
We will not attempt this calculation here and defer it to a future work.
Instead, we shall evaluate the functional traces in two different approximations.
In the first we will neglect all the derivatives of couplings in the r.h.s. of the FRGE
and keep only the explicit dependence of the cutoff on $k$.
In particular we set $\zeta_J=0$, $\zeta_K=0$ in \eqref{dcut}. 
We will refer to this as the {\it one-loop approximation},
since this is the result one would obtain by
inserting a cutoff in the one-loop determinants and then deriving with respect to $k$.
We will see that even though we are tracking the flow of terms with two derivatives of the
background field, this approximation is very similar to the LPA in the linear formalism.
The second approximation is to replace the genuine anomalous dimensions $\eta_J$ and $\eta_K$
by $-\zeta_J$ and $-\zeta_K$, as already indicated in \eqref{dcut}.
We will call this the {\it \sfa}.
In the vicinity of the fixed points that we shall study here the anomalous dimensions
are small anyway and both approximations should be good.
We shall see a posteriori that in spite of the non-vanishing anomalous dimensions
also this approximation is still quite close to the LPA.

In Appendix A we derive explicit formulae (\ref{BetaJ},\ref{BetaK},\ref{BetaV})
for $\zeta_J$, $\zeta_K$, $\zeta_V$,
which are essentially the beta functions for $J$, $K$ and $V$.
At this point we switch to dimensionless variables and simultaneously implement the
coordinate transformation \eqref{coordtransf}.
\footnote{This is equivalent to performing, after each functional integration
over an infinitesimal momentum shell, a rescaling of momenta to restore
the condition $k=1$ and a redefinition of the field to restore its canonical normalization.}
We define $\tilde K=k^{2-d}K$, $\tilde V=k^{-d}V$,
both regarded as functions of the dimensionless field $\tilde h=k^\frac{2-d}{2}h$.
The functions $\zeta_J$, $\zeta_K$, $\zeta_V$, being dimensionless, can
be written in term of the dimensionless variables simply setting $k=1$, $J=1$ and
putting a tilde on $K$ and $V$.

The beta functions can be presented most compactly as follows.
Define the quantities
\bea
\zeta_1 = \frac{c_d}{d+2}\,\frac{N-1}{\tilde{V}(1+\tilde{V}'\tilde{K}'/2\tilde{K})}, \quad \zeta_2 = \frac{c_d}{d+2}&&\!\!\!\!\!\!
\frac{1}{\tilde{V}(1+\tilde{V}'')}, \quad \zeta_3 = \frac{c_d}{d+2}\,\frac{(N-2)(4-\tilde{K}^{\prime 2}/\tilde{K})}{2\tilde{K}(1+\tilde{V}'\tilde{K}'/2\tilde{K})^2}
\nonumber
\\
\zeta_4 = \frac{c_d}{d+2}\,\frac{\tilde{K}^{\prime 2}/\tilde{K}-2\tilde{K}''}{2\tilde{K}(1+\tilde{V}'')^2},
\quad
&&
\zeta_5 = \frac{c_d}{d+2}\, \frac{(N-1)(\tilde{K}^{\prime 2}/\tilde{K}-2\tilde{K}'')}{2\tilde{K}(1+\tilde{V}'\tilde{K}'/2\tilde{K})^2}.
\eea
where $c_d= \frac{1}{(4\pi)^{d/2}\Gamma(d/2+1)}$.
The $\zeta$'s defined in \eqref{etas} are given, in the one loop approximation, by:
\bea
\label{BetaJ1}
\zeta_J &=&(d+2)\zeta_5
\\
\label{BetaK1}
\zeta_K &=& (d+2)(\zeta_3+\zeta_4)
\\
\label{BetaV1}
\zeta_V &=& (d+2)(\zeta_1 +\zeta_2)
\eea
and in the \sfa, by
\bea
\label{etaJ2}
\zeta_J &=& (d+2)\frac{(1 +\zeta_4)\zeta_5}{1-\zeta_3-\zeta_4 \zeta_5}\ ,
\\
\label{etaK2}
\zeta_K &=& (d+2)\frac{\zeta_3+\zeta_4(1+\zeta_5)}{1-\zeta_3-\zeta_4 \zeta_5}\ ,
\\
\label{etaV2}
\zeta_V &=& (d+2)\frac{\zeta_1(1+\zeta_4)
+\zeta_2(1-\zeta_3+\zeta_5)}{1-\zeta_3-\zeta_4 \zeta_5}\ ,
\eea

From the definition of $\tilde h$ and \eqref{coordtransf} one finds
\be
\frac{d\tilde h}{dt}
=\frac{2-d}{2}\tilde h+\frac{1}{2}\int_0^{\tilde h} d x\,\zeta_J(x)
\ .
\ee
Using the definitions of $\tilde K$ and $\tilde V$ in the relations \eqref{etas},
we arrive at the flow equations
\bea
\label{flowK}
\frac{d\tilde{K}}{dt} &=&
\left(\zeta_K -d+2\right)\tilde{K}
+\frac{d-2}{2}\,\tilde{h}\,\tilde{K}' -\frac{1}{2}\tilde{K}'\int_{0}^{\tilde{h}}\!\!\!\! dx \, \zeta_J (x)\ ,
\\
\label{flowV}
\frac{d\tilde V}{dt} &=&
\left(\zeta_V-d\right)\tilde{V}
+\frac{d-2}{2}\,\tilde{h}\,\tilde{V}' -\frac{1}{2}\tilde{V}'\int_{0}^{\tilde{h}}\!\!\!\! dx \, \zeta_J (x) \ .
\eea
These $t$--derivatives take into account, besides the integration over
fluctuations, also the $t$--dependent field redefinition that is necessary to
maintain the field $h$ canonically normalized.
As expected, the redundant variable $J$ has disappeared from the equations:
$\zeta_J$, $\zeta_K$ and $\zeta_V$ are functions of $\tilde K$ and $\tilde V$ only.
Note that if we think of expanding $J$ in Taylor series,
we have not only normalized the kinetic term but also eliminated infinitely many
redundant interaction terms.

It is interesting to consider first the case $N=1$.
This corresponds to $g_{\alpha\beta}=0$, so the term involving the function $\tilde{K}$ is not present in either side of the FRGE.
There are only two equations which govern the running of $V$ and $J$.
One has $\zeta_J=0$ and after the appropriate field redefinitions eq.(\ref{flowV}) becomes
\be
\label{betaV_not_unique_N=1}
\frac{d\tilde V}{dt} = \frac{c_d}{1+\tilde{V}''(\tilde{h})} -d \, \tilde{V}(\tilde{h}) +\left(\frac{d}{2}-1\right)\tilde{h}\,\tilde{V}'(\tilde{h}).
\ee
This reproduces the well--known flow equation for the potential in the LPA.

If we restrict $J$ to be a constant, it can be viewed as the wave function
renormalization of $h$. Then, one would be tempted to identify $\zeta_J$
with (minus) the anomalous dimension of the field $h$.
However, we must stress that the formulas \eqref{BetaJ1} or \eqref{etaJ2}
for $\zeta_J$ are quite different from the formula for the anomalous dimension
for linear scalar field theories in the LPA':
\be
\eta=c_d\frac{(\tilde V''')^2}{(1+\tilde V'')^4}\ {\rm for}\ N=1\ ;
\qquad
\eta=c_d\frac{2(\tilde h\tilde V''-\tilde V')^2}{\tilde h^2(\tilde h+\tilde h\tilde V''-\tilde V')^2}\ {\rm for}\ N>1\ .
\ee
This is actually to be expected, because they are different quantities.
In order to compute the anomalous dimension $\eta$, to be compared with the preceding formula,
one should compute the two-point function of $\xi^0$. We will not discuss this in the present paper.
We note however that for the Gaussian fixed point $\eta$ is expected to be zero, and so is
$\zeta_J$.
For the fixed points that we shall discuss in this paper we expect our approximations to be
acceptable, but for a quantitatively accurate discussion of the Wilson-Fisher fixed point
one should calculate the anomalous dimension separately.

In the rest of the paper we discuss some scaling solutions
and their infinitesimal deformations.
%%%%%%%%%%%%%%%%%%%%%
\section{The flat (Gaussian) fixed point}
%%%%%%%%%%%%%%%%%%%%%

The choice $K=h^2$ corresponds to the flat metric.
This gives $\zeta_3=\zeta_4=\zeta_5=0$.
Plugging this in eqs.(\ref{BetaJ1},\ref{BetaK1},\ref{BetaV1})
or in eqs. (\ref{etaJ2},\ref{etaK2},\ref{etaV2}) gives
both in one--loop approximation and in \sfa,
\be
\zeta_J=0\ ;\quad
\zeta_K=0\ ;\quad
\zeta_V \! =  \frac{c_d}{\tilde{V}} \left[\frac{N-1}{1+\tilde{V}'(\tilde{h})/\tilde{h}} +\frac{1}{1+\tilde{V}''(\tilde{h})}\right]\ .
\ee
Then eqs.(\ref{flowK},\ref{flowV}) become
\be \label{flow_KV_Gaussian}
\frac{d\tilde K}{dt}=0\ ;\quad
\frac{d\tilde V}{dt} = c_d  \left[\frac{N-1}{1+\tilde{V}'(\tilde{h})/\tilde{h}} +\frac{1}{1+\tilde{V}''(\tilde{h})}\right]-d \tilde{V}(\tilde{h}) + \left(\frac{d}{2}-1\right)\tilde{h} \tilde{V}'(\tilde{h})\ .
\ee
This flow equation for the potential agrees with that of the standard linear theory in the LPA. From \eqref{flow_KV_Gaussian} we see that the fixed point condition for $\tilde{K}(\tilde{h})$ is already satisfied and the fixed point condition for $\tilde{V}(\tilde{h})$ is satisfied by a constant potential
$\tilde{V}(\tilde{h}) = c_d N /d$.
We thus have a fixed point solution
\be
\label{fpflat}
\tilde K_* = \tilde h^2\ , \hspace{1cm} \tilde V_* =   c_d N /d\ .
\ee
This is a Gaussian fixed point, corresponding to a free theory at which, in addition to $O(N)$, the theory is also translation invariant.
It exists in any dimension. In $d=3$ there is also a solution with nontrivial
$\tilde V_*$, corresponding the the Wilson-Fisher fixed point,
but we shall not discuss it in this paper.
One expects that the infinitesimal deformations around the Gaussian fixed point are
characterized by canonical critical exponents.
Let us check this explicitly and obtain the corresponding eigenvectors.

\subsection{Linearised equations at one loop}

Linearising the flow equations \eqref{flowK} and \eqref{flowV},
around the fixed point \eqref{fpflat} one finds
\bea
\label{linearKFlat}
\lambda\,\delta\tilde K &=& \tilde K_* \; \delta \zeta_K
-(d-2)\delta \tilde K
+\frac{d-2}{2}\tilde h \delta\tilde K'
-\frac{1}{2}\tilde K_*'\int_{0}^{\tilde h} \!\!\! dx \; \delta \zeta_J(x)
\\
\label{linearVFlat}
\lambda\,\delta\tilde V &=& \tilde V_* \; \delta \zeta_V
+\frac{d-2}{2}\tilde h \delta\tilde V'
\ ,
\eea
where $\lambda$ are scaling exponents to be determined and
$\delta\zeta_V $, $\delta\zeta_J $ and $\delta\zeta_K $ are given by
\bea
\label{deltazetaJ}
\delta \zeta_J\, &=& -\frac{2(N-1)c_d}{\tilde h^4}  \,\delta\tilde K
+\frac{2 (N-1) c_d}{\tilde h^3} \,\delta\tilde K'
-\frac{(N-1) c_d}{\tilde h^2} \,\delta\tilde K''
\\
\label{deltazetaK}
\delta \zeta_K &=& \frac{2 (N-3) c_d}{\tilde h^4}\,\delta\tilde K
-\frac{2 (N-3) c_d}{\tilde h^3}\,\delta\tilde K'
-\frac{c_d}{\tilde h^2} \,\delta\tilde K''
\\
\delta \zeta_V &=& -\frac{d^2}{N c_d} \,\delta\tilde V
-\frac{d (N-1)}{N\tilde h} \,\delta\tilde V'
-\frac{d}{N}\,\delta\tilde V'' %\nonumber
\eea

Eq.\eqref{linearKFlat}, contains the second derivative $\delta\tilde K''$
and an integral. So this is in principle a third order equation.
To get rid of one derivative we divide the equation by
$\tilde K_*'=2\tilde h$, and take its derivative with respect to $\tilde h$.
This gives
\be
\label{linearIFlatDer}
\lambda\; \left(\frac{\delta\tilde K}{\tilde K_*'}\right)' =
\left(\frac{K_*}{\tilde K_*'} \; \delta \zeta_K \right)'
-(d-2)\left(\frac{\delta\tilde K}{\tilde K_*'}\right)'
+\frac{d-2}{4}\delta\tilde K''
-\frac{1}{2}\,\delta \zeta_J
\ee
if we define $\Delta K \equiv \delta\tilde K/\tilde K_*'$ we have
\bea
\label{DeltazetaJ}
\delta \zeta_J\, &=&
-\frac{2(N-1)c_d}{\tilde h}  \,\Delta\tilde K''
\\
\label{DeltazetaI}
\delta \zeta_{K} &=&
-\frac{4(N-2)c_d}{\tilde h^2}\,\Delta\tilde K'
-\frac{2c_d}{\tilde h} \,\Delta\tilde K''
\\
\label{DeltazetaV1loop}
\delta \zeta_V &=& -\frac{d^2}{N c_d} \,\delta\tilde V
-\frac{d (N-1)}{\tilde h N} \,\delta\tilde V'
-\frac{d}{N}\,\delta\tilde V''
\eea
Then we observe that only derivatives of $\Delta\tilde K$ appear in these expressions
and not $\Delta\tilde K$ itself.

Therefore, defining the variable $\dk \equiv (\Delta\tilde K)'$,
the first derivative of eq.(\ref{linearKFlat})
becomes the following second order equation:
\be
\label{lineardiffIpFlat1loop}
0 =\dk''
+\left(\frac{N-3}{\tilde h}
-\frac{d-2}{2c_d}\tilde h\right)\,\dk'
-\left(\frac{2(N-2)}{\tilde h^2}-\frac{\lambda}{c_d}\right)\,\dk\ .
\ee
whereas eq.(\ref{linearVFlat}) becomes
\be
\label{lineardiffVpFlat1loop}
0 = \delta\tilde V''+\frac{2c_d(N-1)-(d-2)\tilde h^2}{2c_d \tilde h}\,\delta\tilde V'
+\frac{d+\lambda}{c_d}\,\delta\tilde V
\ee

Equation (\ref{lineardiffVpFlat1loop}) is well-known in the literature
(see e.g. eq.(A.3) of \cite{litim1}).
Imposing regularity in the origin and boundedness by polynomials for large field leads to
\be
\label{VFlatI0Poly1loop}
\delta\tilde V = \;{}_{1\!}F_1 (-i,N/2,\bar{h}^2), \hspace{5mm} \lambda = (d-2)i-d, \hspace{2mm}i=0, 1, 2, \ldots
\ee
where
\be
\label{barh}
\bar{h} = \left(\frac{d-2}{4c_d}\right)^{\!\! 1/2} \!\! \tilde h.
\ee
In \cite{litim1}, the eigenfunctions were represented in terms of Laguerre polynomials.
In fact our result (\ref{VFlatI0Poly1loop}) agrees with \cite{litim1} noticing that
\be
L^{N/2-1}_n(z)=\frac{(N/2)_n}{n!}\;{}_{1\!}F_1 (-n,N/2,z), \hspace{1cm} (\alpha)_n \equiv \alpha (\alpha +1)\cdots (\alpha +n-1), \;\;\; (\alpha)_0 \equiv 1,
\ee

Similarly, the regular and polynomial--bounded solutions to eq.(\ref{lineardiffIpFlat1loop}) are
\be
\label{IpFlatPoly1loop}
\dk = \bar{h}^2 {}_{1\!}F_1 (-i+1,1+N/2,\bar{h}^2), \hspace{5mm} \lambda = (d-2)i, \hspace{2mm} i=1,2,3, \ldots
\ee
Again, these hypergeometric functions are polynomials.
Note that there is degeneracy with the solutions \eqref{VFlatI0Poly1loop}
when $d/(d-2)$ is an integer. The only dimensions where this happens are $d=0,1,3,4$.
We can convert this to solutions for $\delta\tilde K$
by using the definition $\dk=(\delta\tilde K/\tilde K_*')'$
\be
\delta\tilde K = 2\tilde h
\int_{0}^{\tilde h}\!\!\! dx \,x^2{}_{1\!}F_1 (-i+1,1+N/2,\bar{x}^2)
\hspace{5mm} \bar{x} = \left(\frac{d-2}{4c_d}\right)^{\!\! 1/2} \!\!\!\!\! x.
\ee
The integration constant has been put to zero using eq.(\ref{linearKFlat}).
These solutions have $\dk\not=0$.
There is also a solution $\delta\tilde K \propto \tilde h$ (and hence $\dk=0$)
with eigenvalue $\lambda=1-d/2$.
The only dimension for which the equation for $\delta V$
also has a solution with $\lambda=1-d/2$ is $d=6$.

\subsection{Linearised equations in the \sfa}

In the \sfa\ the linearized equations have the same form
except for the replacement of \eqref{DeltazetaV1loop} by
\bea
\delta \zeta_V &=& -\frac{d^2}{N c_d} \,\delta\tilde V
-\frac{d (N-1)}{N\tilde h} \,\delta\tilde V'
-\frac{d}{N}\,\delta\tilde V'' \nonumber
\\
&&
\label{deltazetaV}
+\frac{2 d (N-1)(N-4) c_d}{N(d+2) \tilde h^4} \,\delta\tilde K
-\frac{2 d (N-1)(N-4) c_d}{N(d+2) \tilde h^3} \,\delta\tilde K'
-\frac{2 d (N-1) c_d}{N(d+2) \tilde h^2}\,\delta\tilde K''
\eea
The terms in the second line introduce a coupling between the
equations for $\delta K$ and $\delta V$.
Proceeding as in the previous section, one arrives at a system of equations
for $\dk$ and $\delta\tilde V$, where \eqref{lineardiffIpFlat1loop} is unchanged and
\eqref{lineardiffVpFlat1loop} is replaced by
\be
\label{lineardiffVpFlat}
0= \delta\tilde V''+\frac{2c_d(N-1)-(d-2)\tilde h^2}{2c_d \tilde h}\,\delta\tilde V'
+\frac{d+\lambda}{c_d}\,\delta\tilde V
+\frac{4(N-1)c_d}{(d+2)\tilde h}\,
\left[\dk'
+\frac{N-2}{\tilde h}\, \dk\right]
\ee
We note that \eqref{lineardiffIpFlat1loop} is automatically satisfied when $\delta\tilde K=0$.
Therefore, the system of equations has an infinite set of solutions where $\delta K=0$
and $\delta V$ is as in the preceding section.
Next we look for solutions to eqs.(\ref{lineardiffVpFlat}, \ref{lineardiffIpFlat1loop}) with
$\dk\neq 0$.
To this end we plug the eigenfunction \eqref{IpFlatPoly1loop},
in eq.(\ref{lineardiffVpFlat}), and solve for $\delta\tilde V$.
The most general solution to this equation consists of any of its solutions
plus the solution to the homogeneous equation \eqref{lineardiffVpFlat1loop}.
Since \eqref{IpFlatPoly1loop} is a polynomial of order $2i$, the two terms in
equation \eqref{lineardiffVpFlat} involving $\dk$ are of order $2(i-1)$.
Therefore a solution of the inhomogeneous equations (\ref{lineardiffVpFlat})
with $\dk$ given by \eqref{IpFlatPoly1loop} can be found by making a general
order $2(i-1)$ polynomial ansatz for $\delta\tilde V$ and solving
for the unknown coefficients.
The solution to the homogeneous equation which is well behaved at $\tilde h=0$ is given by (\ref{VFlatI0Poly1loop}), but now the eigenvalues are found from eq.(\ref{IpFlatPoly1loop}),
so these solutions can be written as
\be
\label{VFlatHomPoly}
\delta\tilde V = {}_{1\!}F_1 (-i-d/(d-2),N/2,\bar{h}^2), \hspace{5mm} \lambda = (d-2)i, \hspace{2mm}i=1, 2, \ldots, \hspace{5mm} \bar{h} = \left(\frac{d-2}{4c_d}\right)^{\!\! 1/2} \!\!\!\!\! h.
\ee

%%%%%%%%%%%%%%%%%%%%%%%%%%%%%%%%%%%%%%%%%%
\section{The cylindrical fixed point}
%%%%%%%%%%%%%%%%%%%%%%%%%%%%%%%%%%%%%%%%%%%%

\subsection{The cylindrical fixed point in the one--loop approximation}

Let us now see if there is a fixed point with $\tilde{K}=\mathrm{const}$.
From eqs.(\ref{BetaJ1},\ref{BetaK1},\ref{BetaV1}) one has
\bea
\label{BetaJ_IVconst1loop}
\zeta_J &=&  0\\
\label{BetaI_IVconst1loop}
\zeta_K \! &=&  2c_d \frac{N-2}{\tilde{K}}
\\
\label{BetaV_IVconst1loop}
\zeta_V \! &=&  c_d \left[\frac{N-1}{\tilde{V}} +\frac{1}{\tilde{V}(1+\tilde{V}'')}\right]
\eea
Using $\zeta_J =0$ and eq.(\ref{flowK}) the fixed point condition implies $\zeta_K=d-2$.
Plugging this back into eq.(\ref{BetaI_IVconst1loop}) one obtains the fixed point value of
$\tilde K= c_d\,\frac{N-2}{d-2}$.
Also combining eq.\eqref{BetaV_IVconst1loop} with eq.\eqref{flowV} gives the fixed point condition
\be
\label{BetaV_IVconst_2}
0= c_d \left[N-1 +\frac{1}{1+\tilde{V}''}\right]-d\tilde{V} +\left(\frac{d}{2}-1\right)h \tilde{V}'
\ee
which is a non-linear second order differential equation.
This equation has a solution for a constant $\tilde V$.
Then the fixed point is
\be \label{V_fp_cyl}
\tilde{K}_* = c_d\,\frac{N-2}{d-2}\ ;\qquad
\tilde{V}_* = \frac{c_d N}{d}.
\ee
We will not consider more general solutions in this work. However it is worth mentioning that we have verified numerically, following the same method used for example in \cite{cododo}, that in $d=4$ this is the only scaling solution while in $d=3$ there also exists a fixed point with nontrivial potential.
The fixed point \eqref{V_fp_cyl} has the geometry of a cylinder $R\times S^{N-1}$.
So at the fixed point the symmetry is enhanced:
in addition to $O(N)$ it is also invariant under translations
in the $\tilde h$ direction.

From the flow equations (\ref{flowK}) and (\ref{flowV}), the linearised equations around the fixed point with constant potential described above have the same form as (\ref{linearKFlat},\ref{linearVFlat}), except that now $\tilde K_*'=0$, so:
\bea
\label{linearICylinder1loop}
\lambda\; \delta\tilde K &=& \tilde K_* \; \delta \zeta_K
-(d-2)\delta\tilde K
+\frac{d-2}{2}\tilde h \delta\tilde K'
\\
\label{linearVCylinder1loop}
\lambda\; \delta\tilde V &=& \tilde V_* \; \delta \zeta_V
+\frac{d-2}{2}\tilde h \delta\tilde V'
\eea
In the above equations, $\delta\zeta_J$, $\delta\zeta_K$ and $\delta\zeta_V$ are given by:
\bea
\label{deltazetaJ}
\delta \zeta_J\, &=&  -\frac{(d-2) (N-1)}{2 (N-2)} \,\delta\tilde K''
\\
\label{deltazetaK}
\delta \zeta_{K} &=& -\frac{(d-2)^2}{2 (N-2) c_d}\,\delta\tilde K -\frac{d-2}{2 (N-2)} \,\delta\tilde K''
\\
\delta \zeta_V &=& -\frac{d^2}{N c_d}\,\delta\tilde V -\frac{d}{N}\,\delta\tilde V''
\label{deltazetaV}
\eea
The second order equations \eqref{linearICylinder1loop} and \eqref{linearVCylinder1loop}
can now be written as the following decoupled system
{\setlength\arraycolsep{2.2pt}
\bea
\label{lineardiffICyl}
0 &=& c_d \,\delta\tilde K''
-\frac{d-2}{2}\tilde h\,\delta\tilde K'
+\left(d-2+\lambda\right)\delta\tilde K\ ,
\\[4mm]
\label{lineardiffVCyl}
0 &=& c_d \,\delta\tilde V''
-\frac{d-2}{2}\tilde h \,\delta\tilde V'
+(d+\lambda)\,\delta\tilde V
\ .
\eea}
Except for the coefficients of $\delta\tilde K$ and $\delta\tilde V$,
these are essentially the same equation.
As in the flat case, regularity constrains the eigenvalue $\lambda$ to discrete values.
The solutions are
\bea \label{deltaV_cyl_1loop}
\delta\tilde V &=& {}_{1\!}F_1 (-i,1/2,\bar{h}^2), \hspace{5mm} \lambda = -d+(d-2)i, \hspace{2mm}i=0, 1, 2, \ldots, \\ \label{deltaK_cyl_1loop}
\delta\tilde K &=& {}_{1\!}F_1 (-i,1/2,\bar{h}^2), \hspace{5mm} \lambda = -d+2+(d-2)i, \hspace{2mm}i=0, 1, 2, \ldots
\eea
Common eigenvalues exist only when $ (d-2)i=2 $ for integer values of $i$, which means when $ 2/(d-2) $ is an integer, or $d= 0, 1, 3, 4$.
%$ d= 2+2/i, \;\; i \in \mathbb{Z}$ .

\subsection{The cylindrical fixed point in the \sfa}

Now we look for a fixed point with $\tilde{K}=\mathrm{const}$
in the \sfa.
From eqs.(\ref{etaJ2},\ref{etaK2},\ref{etaV2}) one has
\bea
\label{BetaJ_IVconst}
\zeta_J &=&  0\\
\label{BetaI_IVconst}
\zeta_K\! &=&  c_d \left(1+\frac{\zeta_K}{d+2}\right)2\frac{N-2}{\tilde{K}}
\\
\label{BetaV_IVconst}
\zeta_V\! &=&  c_d \left[\left(1+\frac{\zeta_K}{d+2}\right)\frac{N-1}{\tilde{V}} +\frac{1}{\tilde{V}(1+\tilde{V}'')}\right]
\eea
Using $\zeta_J =0$, eq.(\ref{flowK}) also gives $\zeta_K= d-2$.
Plugging this back into eqs.(\ref{BetaV_IVconst},\ref{BetaI_IVconst}) gives
\be
\label{ZetaV_IVconst_2}
\tilde{K} = c_d\,\frac{4d \,(N-2)}{d^2-4}\ ;\qquad
\zeta_V \,\tilde{V} =  c_d \left[\frac{2\,d\,(N-1)}{d+2} +\frac{1}{1+\tilde{V}''}\right]\ .
\ee
The first equation gives the fixed point value of the constant function $\tilde{K}$.
Combining the second equation with eq.\eqref{flowV} gives the fixed point condition
\be
\label{BetaV_IVconst_2}
0= c_d \left[\frac{2\,d\,(N-1)}{d+2} +\frac{1}{1+\tilde{V}''}\right]-d\tilde{V} +\left(\frac{d}{2}-1\right)h \tilde{V}'
\ee
This equation has a solution for $\tilde{V}$ constant.
The fixed point is
\be
\tilde{K}_* = c_d\,\frac{4d \,(N-2)}{d^2-4}\ ;\qquad
\tilde{V}_* = c_d \,\frac{2\,d\,(N-1)+d+2}{d(d+2)}\ .
\ee
It is shifted relative to the one-loop solution (\ref{V_fp_cyl}),
but has the same general properties.

The linearised equations around the fixed point have the same form as in
(\ref{linearICylinder1loop},\ref{linearVCylinder1loop}), but now
\bea
\label{deltazetaJ}
\delta \zeta_J\, &=&  -\frac{(d-2)(N-1)}{2(N-2)} \,\delta\tilde K''
\\
\label{deltazetaK}
\delta \zeta_{K} &=& -\frac{(d-2)^2}{2(N-2)c_d}\,\delta\tilde K
-\frac{d-2}{2 (N-2)} \,\delta\tilde K''
\\
\delta \zeta_V &=& -\frac{d^2(d+2)}{(d(2N-1)+2)c_d}\,\delta\tilde V
-\frac{d(d+2)}{d(2N-1)+2}\,\delta\tilde V'' \nonumber
\\
&&
\label{deltazetaV}
-\frac{(d-2)^2 d(N-1)}{2(N-2)(d(2N-1)+2)c_d} \,\delta\tilde K
-\frac{(d-2)d(N-1)}{(N-2)(d(2N-1)+2)}\,\delta\tilde K''
\eea
The two second order equations \eqref{linearICylinder1loop} and \eqref{linearVCylinder1loop}
can now be written as
{\setlength\arraycolsep{2.2pt}
\bea
\label{lineardiffICyl}
0 &=& 4dc_d \,\delta\tilde K''
-\left(d^2-4\right)\tilde h\,\delta\tilde K'
+2\left(2d^2+d(\lambda-4)+2\lambda\right)\delta\tilde K\ ,
\\[4mm]
\label{lineardiffVCyl}
0 &=& c_d \,\delta\tilde V''
-\frac{d-2}{2}\tilde h \,\delta\tilde V'
+(d+\lambda)\,\delta\tilde V
+\frac{(d-2) (N-1) c_d}{(d+2) (N-2)} \,\delta\tilde K''
+\frac{(d-2)^2 (N-1)}{2 (d+2) (N-2)} \,\delta\tilde K\ .
\eea}
To solve this eigenvalue problem, we start again by restricting ourselves to the case
$\delta\tilde K =0$.
In this case the first equation above is automatically satisfied and the second becomes
\be
\label{lineardiffVCylI0}
c_d \,\delta\tilde V''
-\left(d/2-1\right)h \,\delta\tilde V'
+(d+\lambda)\,\delta\tilde V=0.
\ee
This was solved in the preceding section where we found the solution \eqref{deltaV_cyl_1loop}.

Now we look for solutions to eqs.(\ref{lineardiffICyl},\ref{lineardiffVCyl}) with
$\delta\tilde K\neq 0$.
The general regular solution to (\ref{lineardiffICyl}) is
\be
\label{ICylPoly}
\delta\tilde K = {}_{1\!}F_1 (-i,1/2,\bar{h}^2), \hspace{5mm} \lambda = (d-2)i - \frac{2d(d-2)}{d+2}, \hspace{2mm} i=0,1,2,3, \ldots
\ee
We can now plug this solution into eq.(\ref{lineardiffVCyl}), and solve for $ \delta\tilde V $.
The most general solution to this equation consists of any of its solutions plus
a solution to the homogeneous equation eq.(\ref{lineardiffVCylI0}).
A solution of (\ref{lineardiffVCyl}) can be found by making a polynomial ansatz
of order $2i$ and solving for its coefficients.
The solutions to the homogeneous equation which are well behaved at $\tilde h=0$ are
\be
\label{VCylHomPoly}
\delta\tilde V = {}_{1\!}F_1 \left(-i+\frac{d(3d-2)}{d^2-4},1/2,\bar{h}^2\right), \hspace{4mm} \lambda = (d-2)i - \frac{2d(d-2)}{d+2}, \hspace{1mm}i=0, 1, 2, \ldots
\ee
This is a polynomial with a finite number of terms only when $i-\frac{d(3d-2)}{d^2-4}$ is a non-negative integer. This happens only for $d=6$.

%%%%%%%%%%%%%%%%%%%%%%%%%%%%%%%%%%%%%%%%%
\section{The spherical fixed point}
%%%%%%%%%%%%%%%%%%%%%%%%%%%%%%%%%%%%%%%%%

For $J=1$, $K=f^2\sin^2(h/f)$ and $V=\mathrm{const}$,
the symmetry of the Lagrangian is enhanced to $SO(N+1)$.
The quantization procedure used here preserves global symmetries \cite{zanusso},
and since $SO(N+1)$ symmetry fixes completely the form of the metric,
up to the overall factor $f^2$,
and the constant value of the potential, only these two parameters can flow.
A constant potential automatically satisfies $\frac{d\tilde V}{dt}=0$,
and it is known from \cite{codello1} that the flow equation for $\tilde f^2$
has a fixed point.
Therefore it is already clear that there is a fixed point
corresponding to the geometry of $S^N$.
We will nevertheless check this with our equations.
In terms of the rescaled dimensionless variables we have
$\tilde{K}= \tilde{f}^2\sin^2 (\tilde{h}/\tilde{f})$
and $\tilde{V}$ a real constant.
Keeping $\tilde{h}$ fixed, the derivative of $\tilde{K}$ with respect to the scale will be
\be
\label{Beta_I_N+1}
\partial_t\big|_{\tilde{h}} \tilde{K} =
2\partial_t{\tilde{f}}/\tilde{f}\left[\tilde{K}-(\tilde{h}/2)\tilde{K}'\right]
\ee
We discuss first the one loop approximation.

\subsection{The spherical fixed point in the one-loop approximation}

For $\tilde{V}(\tilde{h})=\mathrm{const}$ and
$\tilde{K}(\tilde{h}) = \tilde{f}^2\sin^2 \tilde{h}/\tilde{f}$, eqs.(\ref{BetaJ},\ref{BetaK},\ref{BetaV}) are
\be
\label{Beta_N+1_1loop}
\zeta_J = \zeta_K=2 c_d \frac{N-1}{\tilde{f}^2}\equiv\eta\ ;\qquad
\zeta_V \! =  c_d \frac{N}{\tilde{V}}\ ,
\ee
Eqs.(\ref{flowK},\ref{flowV}) then give
\bea
\label{betaI_not_unique_N+1_1loop}
\frac{d{\tilde{K}}}{dt} &=&
\left(\tilde{K}-(\tilde{h}/2)\tilde{K}'\right)\, \left(\eta -d+2\right)\ ,
\\
\label{betaV_not_unique_N+1_1loop}
\frac{d{\tilde{V}}}{dt}
 &=& \tilde{V}\, \left(\zeta_V -d\right) = c_dN-d\,\tilde{V}\ .
\eea
The first equation leads to
\be
\label{fdot}
2\frac{d{\tilde{f}}}{dt}=(\eta-d+2)\tilde f\ .
\ee
and using \eqref{Beta_N+1_1loop} this reproduces the one-loop result of \cite{codello1}.
The fixed point values are found to be
$\tilde f^2_* = \frac{2c_d (N-1)}{d-2}$, $\tilde{V}_* = \frac{c_d N}{d}$.
Summarizing, the spherical fixed point is given by
\be
\label{V*I*Sphere}
\tilde V_*=\frac{c_d N}{d}, \hspace{1cm}
\tilde K_*=\frac{2c_d (N-1)}{d-2}\,\sin^2\!\bar h,\hspace{1cm} \bar h=\sqrt{\frac{d-2}{2c_d (N-1)}}\, \tilde h.
\ee
In the argument of $\sin$ an arbitrary additive constant has been set to zero.
We have $0\leq \bar h\leq\pi$ which ensures regularity of the metric at $\bar h=0$, $\bar h=\pi$.

The linearised flow equations are
\bea
\label{linearKSph}
\lambda\; \delta\tilde K &=&\tilde K_* \; \delta \zeta_K -\frac{1}{2}\,\tilde K'_*\!\int_{0}^{\tilde h} \!\!\! dx \; \delta \zeta_J (x)
\\
\label{linearVSph}
\lambda\; \delta\tilde V &=&\tilde V_* \; \delta \zeta_V
\eea
It is convenient at this point to use $\bar h$ instead of $\tilde h$
as the argument of all functions and define $\tilde f_*^2 \bar K(\bar h)=\tilde K(\tilde h)$,
$\delta\bar K(\bar h)=\delta\tilde K(\tilde h)$ and $\delta\bar V(\bar h)=\delta\tilde V(\tilde h)$.
Note that a prime on a bar--function differs by a factor $\tilde f_*$
from a prime on a tilde--function. In terms of these bar--functions, $\delta \zeta_J$, $\delta \zeta_K$ and $\delta \zeta_V$ are given by
\bea
\label{deltazetaJsph}
\tilde f_*^2\delta \zeta_J\, &=& -(d-2)\left[
2\cot (\bar h)\,\delta\bar V'
+\csc^4(\bar h)\,\delta\bar K -\cot (\bar h) \csc^2(\bar h)\,\delta\bar K'
+\frac{1}{2}\csc^2(\bar h)\,\delta\bar K''\right]
\\
\label{deltazetaKsph}
\tilde f_*^2\delta \zeta_K &=& -\frac{d-2}{N-1}\left[2(N-2)\cot(\bar h)\,\delta\bar V'+2\,\delta\bar V'' -\csc ^4(\bar h)((N-2) \cos(2\bar h)-1)\,\delta\bar K \right.
\\
&& \left.
+(N-3)\cot(\bar h)\csc^2(\bar h)\,\delta\bar K'
+\frac{\csc^2(h)}{2}\,\delta\bar K'' \right]
\\
\tilde V_*\delta \zeta_V &=& -d\, \delta \bar V -\frac{d-2}{2}\cot (\bar h)\,\delta\bar V'-\frac{d-2}{2 (N-1)}\,\delta\bar V''
\label{deltazetaVsph}
\nn
\eea
Furthermore, in order to reduce the linearized equations to second order,
we proceed as in the flat  case.
Re-expressing $\delta\bar K$ in terms of
$\Delta\bar K=\delta\bar K/\bar K_*'$,
we have
{\setlength\arraycolsep{1pt}
\bea
\label{DeltazetaJsph}
\tilde f_*^2\delta \zeta_J\, &=& -(d-2)\left[
2\cot (\bar h)\,\delta\bar V' -2\,\Delta\bar K'
+\cot(\bar h)\,\Delta\bar K''\right]
\\
\label{DeltazetaKsph}
\tilde f_*^2\delta \zeta_K &=& -\frac{d-2}{N-1}\left[2(N\! -\! 2)\cot(\bar h)\,\delta\bar V'\! + \! 2\,\delta\bar V'' \! +\! 2 \left((N\! -\! 2)\cot^2(\bar h)\! -\! 1\right)\Delta\bar K'
\! +\! \cot(\bar h)\,\Delta\bar K'' \right]
\\
\tilde V_*\delta \zeta_V &=& -d\, \delta \bar V -\frac{d-2}{2}\cot (\bar h)\,\delta\bar V'-\frac{d-2}{2 (N-1)}\,\delta\bar V''
\label{DeltazetaVsph}
\eea}%
Note that there are no undifferentiated $\Delta\bar K$ in these expressions,
and the coefficient of the undifferentiated $\delta\bar V$ is constant.
Therefore, choosing to work with the variables
$\dv \equiv \delta\bar V'$, $\dk \equiv (\Delta\bar K)'$,
the first derivatives of eqs.(\ref{linearKSph},\ref{linearVSph})
become the following second order equations: 

\bea 
\label{lineardiffVSph1loop}
0 &=&  \dv''
+(N-1)\cot(\bar h)\;\dv'
+(N-1)\left(\frac{2(d+\lambda)}{d-2}-\csc^2(\bar h)\right)\dv
\\[4mm]
\label{lineardiffKSph1loop}
0 &=& 2\tan (\bar h)\dv''
+2(\sec^2(\bar h)+N-2)\dv'
-2(N-1)\cot(\bar h)\dv
\\
&& \!\!\!
+\,\dk''
+((N\!-3)\cot(\bar h)-2\tan(\bar h))\dk'
+2\!\left(\frac{(N-1)(d+\lambda-2)}{d-2}-(N-2)\csc^2(\bar h)-\sec^2(\bar h)\right)\dk.
\nn
\eea
As in the preceding two cases, the one-loop linearized equation for $\dv$
contains only $\dv$, but the equation for $\dk$ involves also $\dv$.
Notice that we need not have differentiated eq.\eqref{linearVSph} in which case the equation would have been
\be \label{lineardiffVSph_1loop}
0 =  \delta\bar V''
+(N-1)\cot(\bar h)\;\delta\bar V'
+\frac{2(N-1)(d+\lambda)}{d-2}\delta\bar V\ .
\ee

To define an eigenvalue problem one has to impose some homogeneous boundary conditions.
Assuming regularity of the solutions on the boundaries, equations (\ref{lineardiffVSph1loop},\ref{lineardiffKSph1loop}) force the condition $\dk(0)=\dk(\pi)=0$
and $\dv(0)=\dv(\pi)=0$.
This can be seen by Taylor expanding the equations around these points.

Consider the ``equatorial reflection'' $\bar h\mapsto\pi-\bar h$.
Every solution of eqs.(\ref{lineardiffVSph1loop}, \ref{lineardiffKSph1loop}) has the property that
out of the two functions $\dv$, $\dk$, always one is even and the other is odd.
To see this note that the coefficients of the equations are either even or odd.
Then, defining $\overline\dv(h)=\dv(\pi-h)$ and $\overline\dk(h)=-\dk(\pi-h)$
the functions $\overline\dv$ and $\overline\dk$ satisfy the same equations as $\dv$ and $\dk$.
Since the solutions are unique up to a (common) factor, we must have that
$\dv=\pm\overline{\dv}$ and $\dk=\pm\overline{\dk}$, or
\be \label{parity_vk}
\dv(h)=\pm\dv(\pi-h), \hspace{1cm} \dk(h)=\mp\dk(\pi-h)
\ee
in other words, out of the the pair of functions $\dv$, $\dk$ which satisfy eqs.(\ref{lineardiffVSph}, \ref{lineardiffKSph}), one is even and one is odd.
In the case where $\dk$ is even, $\delta K'$ does not vanish either at $\bar h=0$ or at
$\bar h=\pi$, which means that the metric is singular at least at one endpoint.
For this reason we restrict ourselves to the case when $\dk$ is odd and $\dv$ is even.

We will now describe a method to construct the solutions to eqs.(\ref{lineardiffVSph1loop},\ref{lineardiffKSph1loop}) analytically.
To begin with, we rewrite these equations in the compact form
{\setlength\arraycolsep{3pt}
\bea \label{compact_eqns_K_1loop}
\mathfrak{L}_{\dk\dk} \,\dk + \mathfrak{L}_{\dk\dv} \,\dv &=& \lambda \,\dk \\
\mathfrak{L}_{\dv\dv} \,\dv &=& \lambda \,\dv \label{compact_eqns_V_1loop}
\eea}%
where $\mathfrak{L}_{\dk\dk}$, $\mathfrak{L}_{\dk\dv}$, $\mathfrak{L}_{\dv\dv}$ are differential operators of second order. One can check that
\be
\mathfrak{L}_{\dk\dk}\mathfrak{L}_{\dk\dv}-\mathfrak{L}_{\dk\dv}\mathfrak{L}_{\dv\dv}=2\,\mathfrak{L}_{\dk\dv} \label{operator_id_1}
\ee
This tells us that if $\dv$ is an eigenfunction of $\mathfrak{L}_{\dv\dv}$
with eigenvalue $\lambda$ then $\mathfrak{L}_{\dk\dv}\dv$ is an eigenfunction of
$\mathfrak{L}_{\dk\dk}$ with eigenvalue $\lambda+2$.
Using this we can easily find the solution to eq.(\ref{compact_eqns_K_1loop}). Take $g$ to be a solution to eq.(\ref{compact_eqns_V_1loop}).
Since $\mathfrak{L}_{\dk\dv} g$ is an eigenfunction of $\mathfrak{L}_{\dk\dk}$
we have $\mathfrak{L}_{\dk\dv}g=\eta f$ for some $\eta$ which has to be computed,
where $f$ is an eigenfunction of $\mathfrak{L}_{\dk\dk}$ with eigenvalue $\lambda+2$.
Of course $\eta$ depends on the choice of normalization for $f$ and $g$.
We then plug the ansatz $\dk=C f$ into eq.(\ref{compact_eqns_K_1loop}) to get
\be
(\lambda +2)C f+\eta f = C \lambda f \hspace{5mm} \Rightarrow \hspace{5mm} 2Cf +\eta f =0 \hspace{5mm} \Rightarrow \hspace{5mm} C=-\eta /2
\ee
So with $ g $ given, the function $ -\eta/2 f $ found in this way will be a solution to eq.(\ref{compact_eqns_K_1loop}). So the general solution to eqs.(\ref{compact_eqns_K_1loop},\ref{compact_eqns_V_1loop}) will be $\dv=g$ and $\dk=-\eta/2 f$.
Also if there is a solution to $\mathfrak{L}_{\dk\dk} \,\dk = \lambda \,\dk$ with the same eigenvalue $\lambda$ as that corresponding to $g$ this must also be added to $\dk = -\eta/2 f $. Of course in practice we can solve eq.\eqref{lineardiffVSph_1loop} instead of eq.\eqref{lineardiffVSph1loop}.

For $\delta\tilde V =0$ our equations reduce to $\mathfrak{L}_{\dk\dk} \,\dk = \lambda \,\dk$. A non zero constant $\delta\tilde V$ instead satisfies eq.\eqref{lineardiffVSph_1loop} only if $ \lambda = -d $.

Having found the solutions $ \delta\tilde{V} $ and $\dk$,
the function $ \delta \tilde K(\bar h) $ is of the form
\be
\delta \bar K(\bar h) =  \sin(2\bar h)\int_0^{\bar h} \!\!\! dx \,\dk(x) \ .
\ee
Here an integration constant has been put to zero using equation (\ref{linearKSph}).

\subsection{The spherical fixed point in the \sfa}

In the \sfa\ we find that
for $\tilde{V}(\tilde{h})=\mathrm{const}$ and
$\tilde{K}(\tilde{h}) = \tilde{f}^2\sin^2 \tilde{h}/\tilde{f}$
eqs.(\ref{BetaJ},\ref{BetaK},\ref{BetaV}) are
\bea
\label{BetaJ_N+1}
\zeta_J \, &=& c_d \left(1+\frac{\zeta_K}{d+2}\right)2\frac{N-1}{\tilde{f}^2}\\
\label{BetaI_N+1}
\zeta_K &=& c_d \left[\left(1+\frac{\zeta_K}{d+2}\right)(N-2)+ \left(1+\frac{\zeta_J}{d+2}\right)\right]\frac{2}{\tilde{f}^2}
\\
\label{BetaV_N+1}
\zeta_V  &=&  c_d \left[\left(1+\frac{\zeta_K}{d+2}\right)(N-1) +\left(1+\frac{\zeta_J}{d+2}\right)\right]\frac{1}{\tilde{V}}\ ,
\eea
so also in this case $\zeta_J$, $\zeta_K$ and $\zeta_V$ are constant
and from the first two equations above it is clear that $\zeta_K=\zeta_J\equiv\eta$. Eqs.(\ref{flowK},\ref{flowV}) then give
\bea
\label{betaI_not_unique_N+1}
\frac{d{\tilde{K}}}{dt} &=&
\left(\tilde{K}-(\tilde{h}/2)\tilde{K}'\right)\, \left(\eta -d+2\right)\ ,
\\
\label{betaV_not_unique_N+1}
\frac{d{\tilde{V}}}{dt}
 &=& \tilde{V}\, \left(\zeta_V -d\right) = c_d\left(1+\frac{\eta}{d+2}\right)N-d\,\tilde{V}\ .
\eea
The first equation has the same structure as eq.(\ref{Beta_I_N+1}) and comparing we have
\be
\label{fdot}
2\frac{d{\tilde{f}}}{dt}=(\eta-d+2)\tilde f\ .
\ee
The value of $\eta$ can be found easily from eq.(\ref{BetaJ_N+1}):
\be
\label{zeta}
\eta = -\frac{2c_d (d+2) (N-1)}{2c_d(N-1)-(d+2)\tilde{f}^2}
\ee
This reproduces the result of \cite{codello1}. % eq.(9)
Using \eqref{zeta}, \eqref{betaV_not_unique_N+1} also becomes
\be
\label{Vdot}
\frac{d\tilde V}{dt}= -\frac{c_d (d+2)\tilde{f}^2 N}{2c_d(N-1)-(d+2)\tilde{f}^2}-d\,\tilde{V}.
\ee
From (\ref{fdot},\ref{Vdot}) the fixed point values of $\tilde{f}^2$ and $\tilde{V}$ are found to be
\be
\tilde f^2_* = \frac{4c_d d(N-1)}{d^2-4}\ , \hspace{1cm}
\tilde{V}_* = \frac{2c_d N}{d+2}\ .
\ee
The value of $\tilde V_*$ could also be found more easily from
\eqref{betaV_not_unique_N+1} using the fact that at the fixed point $\eta = d-2$.
So, the spherical fixed point is given by
\be
\label{V*I*Sphere}
\tilde V_* =   \frac{2c_d N}{d+2}, \hspace{1cm}
\tilde K_* =
\frac{4c_d d(N-1)}{d^2-4} \,\sin^2\!\bar h, \hspace{1cm}
\bar h=\sqrt{\frac{d^2-4}{4c_d d(N-1)}} h
\ee
In the argument of $\sin$ an arbitrary additive constant has been set to zero.
we have $0\leq \bar h\leq\pi$ which ensures regularity of the metric at $h=0$, $h=\pi$.

The linearised flow equations have the same structure as eqs.(\ref{linearKSph},\ref{linearVSph}).
We also apply the same definitions explained after eqs.(\ref{linearKSph},\ref{linearVSph}). In this case we have
\bea
\label{deltazetaJsph}
f_*^2\delta \zeta_J &=&
-\frac{8d^2(d-2)(N-1)\cot(\bar h)}{(d+2)(d(2N-1)-2)}\,\delta\bar V'
-\frac{4d(d-2)^2}{(d+2)(d(2N-1)-2)}\,\delta\bar V''
\nn
\\
&&
-\frac{2d(d-2)((d(2N-1)-2)\cos(2\bar h)-3d-4N+10)\csc^2(\bar h)}{(d+2)(d(2N-1)-2)}\,
\Delta\bar K'
\nn
\\
&&
-\frac{2d(d-2)(dN+d+2N-6)\cot(\bar h)}{(d+2)(d(2N-1)-2)}\,
\Delta\bar K''
\\ [4mm]
\label{deltazetaKsph}
f_*^2\delta \zeta_K &=&
-\frac{4d(d-2)(d(2N-3)-2)\cot(\bar h)}{(d+2)(d(2N-1)-2)}\,\delta\bar V'
-\frac{8d^2(d-2)}{(d+2)(d(2N-1)-2)}\,\delta\bar V''
\nn
\\
&&
-\frac{2d(d-2)((d(2N-1)-2)\cos(2\bar h)+d(2N-7)+2)\csc^2(\bar h)}{(d+2)(d(2N-1)-2)}\,
\Delta\bar K'
\nn
\\
&&
-\frac{2d(3d-2)(d-2)\cot(\bar h)}{(d+2)(d(2N-1)-2)}\,\Delta\bar K''
\\[4mm]
\tilde V_*\delta \zeta_V &=& -d\,\delta\bar V
-\frac{(3d-2)(d-2)\cot(\bar h)}{2(d+2)}\,\delta\bar V'
-\frac{(3d-2)(d-2)}{2(d+2)(N-1)}\,\delta\bar V''
\nn
\\
&&
\label{deltazetaVsph}
-\frac{(d-2)^2(N\cos(2\bar h)+N-4)\csc^2(\bar h)}{2(d+2)(N-1)}\,\Delta\bar K'
-\frac{(d-2)^2\cot(\bar h)}{(d+2)(N-1)}\,\Delta\bar K''
\eea
Note that there are no undifferentiated $\Delta\bar K$ in these expressions,
and the coefficient of the undifferentiated $\delta\bar V$ is constant.
Therefore, choosing to work with the variables
$\dv \equiv \delta\bar V'$, $\dk \equiv (\Delta\bar K)'$,
the first derivatives of eqs.(\ref{linearKSph},\ref{linearVSph})
become the following second order equations:
\bea
\label{lineardiffVSph}
0 &=&  \dv''
+(N-1)\cot(\bar h)\;\dv'
+\frac{(N-1)(2(d+2)(d+\lambda)-(3d-2)(d-2)\csc^2(\bar h))}{(3d-2)(d-2)}\;\dv
\\
&+&
\frac{2(d-2)\cot(\bar h)}{3d-2}\;
\left[\dk''
+(N\cos(2\bar h)+N-6)\csc(2\bar h)\;\dk'
-2(N-2)\csc^2(\bar h)\;\dk\right] ,
\nn
\eea
\bea
\label{lineardiffKSph}
0 &=& \frac{4 d \tan (\bar h)}{3 d-2}\;
\left[\dv''
+\left(\sec^2(\bar h)+N-2\right)\cot(\bar h)\;\dv'
-(N-1)\cot(\bar h)^2\;\dv\right]
\\
&+&
\dk''
+((N-1)\cos(2\bar h)+N-5)\csc(2\bar h)\;\dk'
\nn \\
&+&
\frac{(d(2N-1)-2)((d+2)\lambda +2d(d-2))-d(d-2)(3d-2)(2(N-2)\csc^2(\bar h)+2\sec^2(\bar h))}{(d-2)d(3d-2)}\; \dk.
\nn
\eea
Again, assuming regularity of the solutions on the boundaries, equations (\ref{lineardiffVSph},\ref{lineardiffKSph}) force the condition $\dk(0)=\dk(\pi)=0$
and $\dv(0)=\dv(\pi)=0$.
As in the one-loop approximation, we are interested in the solutions where $\dk$
is odd and $\dv$ is even under the equatorial reflection $\bar h\mapsto \pi-\bar h$.

Having found $\dv$, $\dk$,
the solutions to eqs.(\ref{linearKSph},\ref{linearVSph}) are of the form
\be
\delta \bar V(\bar h) =  \int_0^{\bar h} \!\!\! dx \,\dv(x), \hspace{1cm}
\delta \bar K(\bar h) =  \sin(2\bar h)\int_0^{\bar h} \!\!\! dx \,\dk(x)
\ee
Here two integrations constants have been put to zero, as can be deduced from
eqs.(\ref{linearKSph},\ref{linearVSph}) and parity considerations.
In addition to the above solutions with nonvanishing $\kappa$ or $v$,
there are two solutions: $\delta\bar V =C, \,\delta\bar K =0$ with eigenvalue $\lambda=-d$ and $\delta\bar V =0, \,\delta\bar K =C\sin(2\bar h)$ with eigenvalue $\lambda=0$.

In order to solve eqs.(\ref{lineardiffVSph},\ref{lineardiffKSph}) analytically,
we rewrite them in the compact form
{\setlength\arraycolsep{3pt}
\bea
\mathfrak{L}_{\dk\dk} \,\dk + \mathfrak{L}_{\dk\dv} \,\dv &=& \lambda \,\dk \label{compact_eqns_K}\\
\mathfrak{L}_{\dv\dk} \,\dk + \mathfrak{L}_{\dv\dv} \,\dv &=& \lambda \,\dv \label{compact_eqns_V}
\eea}
where $ \mathfrak{L}_{\dk\dk}, \mathfrak{L}_{\dk\dv}, \mathfrak{L}_{\dv\dk}, \mathfrak{L}_{\dv\dv} $ are second order differential operators satisfying
{\setlength\arraycolsep{3pt}
\bea \label{operator_id_1}
\mathfrak{L}_{\dk\dk}\mathfrak{L}_{\dk\dv} -\alpha \,\mathfrak{L}_{\dk\dv}\mathfrak{L}_{\dv\dv} &=& \beta \,\mathfrak{L}_{\dk\dv} \\ \label{operator_id_2}
\mathfrak{L}_{\dv\dk}\mathfrak{L}_{\dk\dk} -\alpha \,\mathfrak{L}_{\dv\dv}\mathfrak{L}_{\dv\dk} &=& \beta \,\mathfrak{L}_{\dv\dk}
\eea}
where
\be
\alpha = \frac{2d(N-1)}{2dN-d-2}, \hspace{1cm} \beta = -\frac{2d(d(d-6)N+2d+4)}{(d+2)(d(2N-1)-2)}
\ee
and also
{\setlength\arraycolsep{3pt}
\bea
\label{operator_id_3}
\mathfrak{L}_{\dk\dk}^2+\gamma_1\,\mathfrak{L}_{\dk\dv}\mathfrak{L}_{\dv\dk}+\gamma_2\,\mathfrak{L}_{\dk\dk}+\gamma_3 1 &=& 0
\\
\label{operator_id_4}
\mathfrak{L}_{\dv\dv}^2+\delta_1\,\mathfrak{L}_{\dv\dk}\mathfrak{L}_{\dk\dv}+\delta_2\,\mathfrak{L}_{\dv\dv}+\delta_3 1 &=& 0
\eea}
where
{\setlength\arraycolsep{3pt}
\be
\ba{lll}
\gamma_1 &=& \displaystyle  \frac{(2-3d)^2(N-1)}{4(d-2)(-2dN+d+2)}        \\[4mm]
\gamma_2 &=& \displaystyle  \frac{(d-2)d(5dN-4d+2N-8)}{(d+2)(d(2N-1)-2)}  \\[4mm]
\gamma_2 &=& \displaystyle  \frac{2(d-2)^2d^2(N-2)}{(d+2)(d(2N-1)-2)}
\ea \hspace{1cm}
\ba{lll}
\delta_1 &=& \displaystyle  -\frac{(2-3d)^2(d(2N-1)-2)}{16(d-2)d^2(N-1)}  \\[4mm]
\delta_2 &=& \displaystyle  -\frac{d(8-16N)-d^2(N-4)+4N}{2(d+2)(N-1)}     \\[4mm]
\delta_2 &=& \displaystyle  -\frac{d(N(d(d-12)+4)+2d(d+2))}{2(d+2)(N-1)}
\ea
\ee}%
Now suppose that $f,g$ are eigenfunctions of $\mathfrak{L}_{\dk\dk},\mathfrak{L}_{\dv\dv}$
with eigenvalues $\sigma$, $\sigma'$:
\be
\label{eigenL}
\mathfrak{L}_{\dk\dk} \,f = \sigma \, f, \hspace*{1cm} \mathfrak{L}_{\dv\dv} \,g = \sigma' \, g\ .
\ee
Then, using the identities (\ref{operator_id_1}) we find that
$\mathfrak{L}_{\dv\dk}f$ is an eigenvector of $\mathfrak{L}_{\dv\dv}$ with eigenvalue $(\sigma - \beta)/\alpha$ and $\mathfrak{L}_{\dk\dv}g$ is an eigenvector of
$\mathfrak{L}_{\dk\dk}$ with eigenvalue $\alpha\sigma'+\beta$
\bea
\mathfrak{L}_{\dv\dv}(\mathfrak{L}_{\dv\dk}\,f) &=&
\frac{\sigma-\beta}{\alpha}\;\mathfrak{L}_{\dv\dk}\,f
\\
\mathfrak{L}_{\dk\dk}(\mathfrak{L}_{\dk\dv} \,g)&=&
\left(\alpha\sigma' + \beta\right)\mathfrak{L}_{\dk\dv}\,g
\eea
Assuming non degeneracy of the eigenvalues of $\mathfrak{L}_{\dk\dk}$, $\mathfrak{L}_{\dv\dv}$,
this means that given $f$, there is a $g$ such that
\be
\mathfrak{L}_{\dv\dk} \,f = \eta \,g, \hspace{1cm} \mathfrak{L}_{\dk\dv} \,g = \eta' \,f
\ee
for some $ \eta, \eta' $. Notice that the transformations
\be
\sigma \rightarrow \frac{\sigma - \beta}{\alpha}, \hspace{1cm} \sigma' \rightarrow \alpha\, \sigma' + \beta
\ee
are inverse of each other, this means that there is a one to one correspondence between the eigenvalues of $ \mathfrak{L}_{\dk\dk} $ and $ \mathfrak{L}_{\dv\dv} $ given by $ \sigma \leftrightarrow \frac{\sigma - \beta}{\alpha} $. From now on we take $ f,g $ to be the eigenvectors which correspond to each other, i.e, the eigenvectors corresponding to the eigenvalues $ \sigma $ and $ \frac{\sigma - \beta}{\alpha} $ respectively.\\
Since we have not fixed the normalization of $ f,g $ we shouldn't expect to be able to find the values of both $ \eta $ and $ \eta' $, because for example making the redefinition $ f\rightarrow \tau f $ leads to $ \eta \rightarrow \tau^{-1}\eta $ and $ \eta' \rightarrow \tau \eta' $. The product $ \eta \eta' $, however, is fixed, and in fact it is an eigenvalue of $ \mathfrak{L}_{\dv\dk}\mathfrak{L}_{\dk\dv}$ and $ \mathfrak{L}_{\dk\dv}\mathfrak{L}_{\dv\dk} $. To find it we exploit one of the identities (\ref{operator_id_3},\ref{operator_id_4}). For example acting on $ f $ by the first equation leads to
\be
\sigma^2 +\gamma_1\,\eta\eta' +\gamma_2\,\sigma +\gamma_3 = 0
\ee
As anticipated, $ \eta $ and $ \eta' $ appear only as $ \eta\eta' $. Solving this equation we find
\be
\eta\eta' = \frac{4(d-2)((d+2)\sigma+2(d-2)d)(\sigma(d(2N-1)-2)+(d-2)d(N-2))}{(2-3d)^2(d+2)(N-1)}.
\ee
Acting on $ g $ by eq.(\ref{operator_id_4}) will lead to the same result.
Let us now plug our two eigenfunctions accompanied by unknown factors $ c_1\, f $ and $ c_2\, g $ into the equations (\ref{compact_eqns_K},\ref{compact_eqns_V}). Doing this we get the linear equation
{\setlength\arraycolsep{3pt}
\be
\label{matrix_eqns_c1c2}
\left(
\ba{cc}
\sigma & \eta' \\
\eta   & \frac{\sigma - \beta}{\alpha}
\ea
\right) \left(\ba{c}c_1 \\ c_2\ea\right) = \lambda \left(\ba{c}c_1 \\ c_2\ea\right)
\ee}%
whose solution gives us two eigenvalues $\lambda^\pm$ and
the corresponding eigenvectors 
$c^{\pm}=(c^\pm_1 \, ,\, c^\pm_2)^T$.
In this way we find two solutions to the equations (\ref{compact_eqns_K},\ref{compact_eqns_V}):
$\dk=c^\pm_1\,f$, $\dv=c^\pm_2\,g$, corresponding to $\lambda=\lambda^\pm$.

It is possible to find explicit expressions in terms of $ d $, $ N $ and $ \sigma $ for the eigenvalues $ \lambda^\pm $ by solving the characteristic equation of the matrix in eq.(\ref{matrix_eqns_c1c2}). This is because it is only the product $ \eta\eta' $ which appears in the characteristic equation. The result in terms of $ \alpha, \beta $ and $ \eta\eta' $ (which all depend on $ d,N $) and $  \sigma $ is
\be
\lambda^\pm = \frac{\sigma(\alpha +1)-\beta \pm\sqrt{4 \alpha^2\eta\eta' +((\alpha -1)\sigma +\beta)^2}}{2\alpha}
\ee
The eigenvectors are
\be
c^\pm_1 = \frac{\sigma(\alpha -1)+\beta \pm\sqrt{4 \alpha^2\eta\eta' +((\alpha -1)\sigma +\beta)^2}}{2\alpha} \hspace{1cm} c^\pm_2 = \eta
\ee
we see in this case that there is a dependence on $\eta$.
After specifying $d$, $N$ one can compute $\eta$ from the formula $\mathfrak{L}_{\dk\dv}\,g=\eta f$.
Having found $\eta$ one can follow the method described above to find analytic solutions for the eigenfunctions. Analytic formulas for the eigen-perturbations in $ d=4 $ and $ N=4 $
are reported in section VI.

%%%%%%%%%%%%%%%%%%%%%%%%%%
\section{The case $d=4$, $N=4$}
%%%%%%%%%%%%%%%%%%%%%%%%%%

Up to this point all formulas hold in arbitrary dimension $d$ and for arbitrary $N$.
As an example we give now the scaling dimensions (eigenvalues of the stability matrix)
and operators (eigenvectors of the stability matrix)
in the special case $d=4$ and $N=4$, which is directly relevant for standard model physics.
With all three topologies there is a relevant perturbation with eigenvalue
$\lambda=-4$ and eigenfunctions $\delta\tilde K=0$, $\delta\tilde V=1$.
This is just a change of the vacuum energy and is only important in a gravitational context.
In fact, we have seen that it is convenient and customary to study the derivative
of the potential, and constants can be neglected.
We will not discuss this mode further in the following.

We begin by considering the Gaussian fixed point.
There is a sequence of perturbations with $\delta\tilde K=0$
with eigenvalues $-2,0,2,4\ldots$;
these are just the canonical dimensions of the couplings
that multiply the operators $\phi^{2n}$, $n=1,2,3,4\ldots$.
Another series of perturbations with nonzero $\delta\tilde K$
have eigenvalues $0,2,4,\ldots$, corresponding to the canonical dimensions of the
couplings that multiply the operators $(\partial\phi)^2\phi^{2n}$, $n=0,1,2\ldots$.
For positive eigenvalues there is therefore a double degeneracy.
The eigenfunctions are not simply given by these field monomials:
they correspond to certain linear combinations.
In the single-field approximation the first few eigenfunctions are
\be
\ba{ll}
\lambda = -2 \hspace{1cm} & \delta\tilde K = 0
             \hspace{1cm} \delta\tilde V = \displaystyle 1-8\pi^2 \tilde h^2
\\[5mm]
\lambda = 0  \hspace{1cm} & \delta\tilde K = 0
             \hspace{1cm}  \delta\tilde V = \displaystyle 1-16\pi^2 \tilde h^2 + \frac{128\pi^4}{3}\tilde h^4 \\[5mm]
\lambda = 2  \hspace{1cm} & \delta\tilde K = \displaystyle \frac{2}{3}\tilde h^4 C_2 \hspace{1cm} \\
             \hspace{1cm} & \delta\tilde V = \displaystyle -\frac{1}{768\pi^4}C_2 + C_1 \left(1-24\pi^2 \tilde h^2 +128\pi^4 \tilde h^4 -\frac{512\pi^6}{3}\tilde h^6\right)
\\[6mm]
\lambda = 4  \hspace{1cm} & \delta\tilde K = \displaystyle C_2\left(\frac{2}{3}\tilde h^4 -\frac{32\pi^2}{15}\tilde h^6\right)            \hspace{1cm}  \\
             \hspace{1cm} & \delta\tilde V = \displaystyle C_2 \left(-\frac{1}{768 \pi ^4}+ \frac{1}{96 \pi ^2}\tilde h^2\right)\\
             \hspace{1cm} & \hspace{6mm} \displaystyle +\, C_1\left(1-32\pi^2 \tilde h^2+256\pi^4 \tilde h^4-\frac{2048\pi^6}{3}\tilde h^6+\frac{8192\pi^8}{15}\tilde h^8\right)
\\[6mm]
\lambda = 6  \hspace{1cm} & \delta\tilde K = \displaystyle C_2\left(\frac{2}{3}\tilde h^4 -\frac{64\pi^2}{15}\tilde h^6+\frac{128\pi^4}{21}\tilde h^8\right) \hspace{1cm}  \\
             \hspace{1cm} & \delta\tilde V = \displaystyle C_2 \left(-\frac{1}{768 \pi ^4}+\frac{1}{48 \pi ^2}\tilde h^2-\frac{1}{18}\tilde h^4\right)\\
             \hspace{1cm} & \hspace{6mm} \displaystyle +\, C_1\left(1-40\pi^2 \tilde h^2+\frac{1280\pi^4}{3}\tilde h^4-\frac{5120\pi^6}{3}\tilde h^6+\frac{8192\pi^8}{3}\tilde h^8-\frac{65536\pi^{10}}{45}\tilde h^{10}\right)  \nn
\ea
\ee
Here $C_1$ and $C_2$ are arbitrary coefficients; there are degenerate eigenfunctions
corresponding to taking $C_1=0$, $C_2\not=0$ or $C_1\not=0$, $C_2=0$.
In the one-loop approximation the equations for $\delta\tilde K$ and $\delta\tilde V$
are independent and the eigenfunctions of type $\delta\tilde V$ can be obtained
from the preceding formulas by setting $\tilde C_2=0$
(in other words, they are the coefficients of $C_1$).

In the case of the cylindrical fixed point in the one loop approximation
the equations for $\delta\tilde K$ and $\delta\tilde V$ are decoupled, so we have independent eigenfunctions for $\delta\tilde K$ and $\delta\tilde V$.
The first few eigenvalues and eigenfunctions are
\be
\ba{ll}
\lambda = -2            \hspace{1cm} & \delta\tilde K = 1
                        \hspace{1cm}  \delta\tilde V = \displaystyle 1-32\pi^2 \tilde h^2
\\[2mm]
\lambda = 0  \hspace{1cm} & \delta\tilde K = \displaystyle 1-32\pi^2 \tilde h^2
             \hspace{1cm} \delta\tilde V = \displaystyle 1-64\pi^2 \tilde h^2 +\frac{1024\pi^4}{3}\tilde h^4
\\[2mm]
\lambda = 2       \hspace{1cm} & \delta\tilde K = \displaystyle \displaystyle 1-64\pi^2 \tilde h^2 + \frac{1024\pi^4}{3} \tilde h^4 \\
                  \hspace{1cm} &
                  \delta\tilde V = \displaystyle 1-96\pi^2 \tilde h^2+1024\pi^4 \tilde h^4 - \frac{32768\pi^6}{15}\tilde h^6
\\[2mm]
\lambda = 4  \hspace{1cm} & \delta\tilde K = \displaystyle 1-96\pi^2 \tilde h^2+1024\pi^4 \tilde h^4 - \frac{32768\pi^6}{15}\tilde h^6 \\
                        \hspace{1cm} & \delta\tilde V = \displaystyle 1-128\pi^2 \tilde h^2+2048\pi^4 \tilde h^4-\frac{131072\pi^6}{15}\tilde h^6+\frac{1048576\pi^6}{105}\tilde h^8
\ea \nn
\ee
In this case we have not written the arbitrary coefficients $C_1$ and $C_2$:
one can take arbitrary linear combinations of the degenerate $\delta\tilde K$ and $\delta\tilde V$.
Note that also in this case the eigenvalues for the potential are just the canonical dimensions
of the operators with the highest power of $h$ minus four, while those for $K$ are
the dimension of the highest power of $h$ minus two (accounting for the extra two derivatives
of the Goldstone bosons).

In the case of the cylindrical fixed point in the \sfa\
there are again two sequences of eigen-perturbations, but they are no longer degenerate.
Those with $\delta\tilde K=0$ and $\delta\tilde V\not=0$ are the same as in the
one--loop approximation, and also have the same eigenvalues.
This is because the field $h$ is just a free field decoupled from the Goldstone bosons.
The eigenfunctions with $\delta\tilde K\not=0$ now must also have $\delta\tilde V\not=0$
and have non-integer eigenvalues that differ by multiples of two.
The eigenfunctions listed below have been obtained by solving the inhomogeneous equation
with the method described in section IVB:
\be
\ba{ll}
\lambda = -\frac{8}{3}  \hspace{1cm} & \delta\tilde K = 1
                        \hspace{1cm}  \delta\tilde V = \displaystyle -\frac{3}{8}
\\[2mm]
\lambda = -\frac{2}{3}  \hspace{1cm} & \delta\tilde K = \displaystyle 1-24\pi^2 \tilde h^2
                        \hspace{1cm}  \delta\tilde V = \displaystyle -\frac{3}{32}+9\pi^2 \tilde h^2
\\[2mm]
\lambda = \frac{4}{3}  \hspace{1cm} & \delta\tilde K = \displaystyle 1-48\pi^2 \tilde h^2 + 192\pi^4 \tilde h^4
\\
                       \hspace{1cm} &
                       \delta\tilde V = \displaystyle \frac{69}{512}+\frac{9\pi^2}{2}\tilde h^2 - 72\pi^4 \tilde h^4
\\[2mm]
\lambda = \frac{10}{3}  \hspace{1cm} & \delta\tilde K = \displaystyle 1-72\pi^2 \tilde h^2+576\pi^4 \tilde h^4 - \frac{4608\pi^6}{5}\tilde h^6 \\
                        \hspace{1cm} & \delta\tilde V = \displaystyle \frac{7239}{22528}-\frac{621\pi^2}{64} \tilde h^2-54\pi^4 \tilde h^4+\frac{1728\pi^6}{5}\tilde h^6
\ea \nn
\ee
The eigenfunction with the most negative eigenvalue ($-2$ at one loop and
$-8/3$ in the \sfa) have already been mentioned in \cite{codello1},
as the slope of the beta function of the sigma model coupling.

Finally let us come to the spherical fixed point. In the one-loop approximation we proceed as explained in section V.A. There will be a series of solution with $\delta\bar V =0$ and $\dk$ satisfying $\mathcal{L}_{\dk\dk}\dk =\lambda \dk$, the (odd) eigenfunctions $f_i$ and corresponding eigenvalues $\sigma_i$ which satisfy this equation are
\be
f_i = \cot(\bar h)\csc(\bar h)\,{}_{2\!}F_1(i+1/2, -i, 5/2, \cos^2\!(\bar h)) \hspace{1cm}\sigma_i = \frac{2}{3} \left(2i^2+i-4\right), \hspace{5mm} i=2,3,\cdots
\ee
where $ {}_{2\!}F_1(a, b, c, z) $ is the hypergeometric function.
For $\delta \bar V \neq 0$ the solution of \eqref{lineardiffVSph_1loop} is
\be
g_i = \csc(h)\, P^1_{2i}(\cos(h)) \hspace{1cm}\sigma'_i = \frac{2}{3} \left(2i^2+i-7\right), \hspace{5mm} i=1,2,\cdots
\ee
where $ P^m_n(x) $ is the associated Legendre polynomial. Now, given $g_i$, the function $-\frac{1}{2}\mathcal{L}_{\dk\dv} g'_i = -\frac{\eta}{2}f_i$ will satisfy eq.\eqref{lineardiffKSph1loop}. Furthermore, there is no solution to eq.\eqref{lineardiffKSph1loop} with $\dv =0$ which has the same eigenvalue as that of $g_i$. So $-\frac{1}{2}\mathcal{L}_{\dk\dv} g'_i = -\frac{\eta}{2}f_i$ will be the unique solution to  eq.\eqref{lineardiffKSph1loop}. The first few eigenvectors and their corresponding eigenvalues are
\be
\ba{ll}
\lambda_1 =\displaystyle -\frac{8}{3}\approx -2.667,  & \hspace{1cm}\delta \bar K = 0\ ;
    \hspace{1cm} \delta \bar V = \displaystyle -3 \cos (\bar h)
\\[3mm]
\lambda_2 =2, & \hspace{1cm}\delta \bar K = \displaystyle -140\sin ^4(\bar h) \cos(\bar h)\ ;  %\\[2mm]
             %& \hspace{1cm}
             \quad\delta \bar V = \displaystyle -\frac{5}{4}\cos(\bar h)(1+7\cos(2\bar h)) \\[3mm]
\lambda_3 = 4, &\hspace{1cm}\delta \bar K = \displaystyle \frac{2}{3}\sin^4(\bar h) \cos(\bar h)\ ;
               \hspace{1cm}\delta \bar V = 0
               \\[3mm]
\lambda_4 =\displaystyle \frac{28}{3}\approx 9.33, &\hspace{1cm}\delta \bar K = \displaystyle -84 \sin ^4(\bar h) \cos (\bar h) (9+11 \cos (2 \bar h)) \\[2mm]
                                                 & \hspace{1cm}\delta \bar V = \displaystyle -\frac{21}{128}(50 \cos (\bar h)+45 \cos (3 \bar h)+33 \cos (5 \bar h))
                                                 \\[3mm]
\lambda_5 =\displaystyle \frac{34}{3}\approx 11.33, & \hspace{1cm}\delta \bar K = \displaystyle -\frac{1}{25} \sin ^4(\bar h) \cos (\bar h) (9+11 \cos (2 \bar h))\ ;% \\[2mm]
                                                  \hspace{1cm}\delta \bar V = 0 \\[3mm]

\lambda_6 =\displaystyle \frac{58}{3}\approx 19.33, & \hspace{1cm}\delta \bar K = \displaystyle -\frac{33}{16} \sin ^4(\bar h) \cos (\bar h) (1581+1924 \cos (2 \bar h)+975 \cos (4 \bar h)) \\[2mm]
& \hspace{1cm}\delta \bar V = \displaystyle -\frac{9}{512} \cos (\bar h) (178+869 \cos (2 \bar h)+286 \cos (4 \bar h)+715 \cos (6 \bar h)) \nn
\ea
\ee

In the \sfa, the solutions of the equations (\ref{eigenL}) are
{\setlength\arraycolsep{3pt}
\bea
\displaystyle f_i &=& \displaystyle \cot(\bar h)\csc(\bar h)\,{}_{2\!}F_1(i+1/2, -i, 5/2, \cos^2\!(\bar h))  \hspace{1.2cm}\sigma_i \,=\, \displaystyle \frac{8}{39} \left(10 i^2+5 i-18\right) \\[3mm]
\displaystyle g_i &=& \displaystyle \csc(h)\, P^2_{2i}(\cos(\bar h)) \hspace{5.12cm}\sigma'_i \,=\, \displaystyle \frac{2}{9} \left(10i^2+5i-23\right)
\eea}%
With these one constructs two series of eigenvectors
$(\delta \tilde V_i^\pm,\delta\tilde K_i^\pm)$ and eigenvalues $\lambda_i^\pm$,
whose first few members are listed below.
\be
\ba{l}
\lambda^-_1 =\displaystyle -\frac{16}{9}\approx -1.778 ;
\hspace{1cm}\delta \bar K_1^- = 0,
\hspace{1cm}\delta \bar V_1^- = \displaystyle \frac{2}{3}\cos \bar h \\[3mm]

\lambda^\pm_2 = \displaystyle \frac{1}{39} \left(245 \pm \sqrt{70009}\right)
\approx (-0.502,13.07)   \\[3mm]
\hspace*{2cm}\delta \bar K_2^\pm = \displaystyle \frac{2(11\pm \sqrt{70009})}{117}\sin ^4(\bar h) \cos (\bar h) \\[2mm]
\hspace*{2cm}\delta \bar V_2^\pm = \displaystyle -\frac{1}{18} (9 \cos (\bar h)+7 \cos (3 \bar h)) \\[3mm]
\lambda^\pm_3 =\displaystyle \frac{2}{117} \left(1055\pm \sqrt{898681}\right)
\approx(1.829,34.24)   \\[3mm]
\hspace*{2cm}\delta \bar K_3^\pm = \displaystyle \frac{2 \left(11\mp\sqrt{898681}\right)}{2925}\sin ^4(\bar h) \cos (\bar h) (9+11\cos(2\bar h)) \\[2mm]
\hspace*{2cm}\delta \bar V_3^\pm = \displaystyle \frac{1}{120} (50 \cos (\bar h)+45 \cos (3 \bar h)+33 \cos (5 \bar h)) \\[3mm]
\lambda^\pm_4 =\displaystyle \frac{1}{117} \left(3985 \pm \sqrt{11540929}\right)
\approx(5.024,69.10)    \\[3mm]
\hspace*{2cm}\delta \bar K_4^\pm = \displaystyle -\frac{\left(97\mp\sqrt{11540929}\right)}{573300}\sin^4(\bar h)\cos(\bar h)(1581+1924\cos(2\bar h)+975\cos(4\bar h)) \\[2mm]
\hspace*{2cm}\delta \bar V_4^\pm = \displaystyle -\frac{1}{3360}(1225 \cos (\bar h)+11 (105 \cos (3 \bar h)+91 \cos (5 \bar h)+65 \cos (7 \bar h))) \nn %\\[3mm]
\ea
\ee
\be
\ba{l}
\lambda^\pm_5 =\displaystyle \frac{8}{39} \left(265\pm 4\sqrt{3046}\right)
\approx(9.074,99.64) \\[5mm]
\hspace*{2cm}\delta \bar K_5^\pm = \displaystyle \frac{2 \left(2\mp\sqrt{3046}\right)}{51597}\sin ^4(\bar h) \cos (\bar h) (8859 \cos (2 \bar h)\! +\! 4794 \cos (4 \bar h)\! +\! 2261 \cos (6 \bar h)\! +\! 5590) \\[2mm]
\hspace*{2cm}\delta \bar V_5^\pm = \displaystyle \frac{1}{24192}(7938 \cos (\bar h)\! +\! 13 (588 \cos (3 \bar h)\! +\! 540 \cos (5 \bar h)\! +\! 459 \cos (7 \bar h)\! +\! 323 \cos (9 \bar h))) \\[9mm]

\lambda^\pm_6 =\displaystyle \frac{1}{117} \left(9235\pm \sqrt{57751849}\right)
\approx (13.98,143.88) \\[5mm]
\hspace*{2cm}\delta \bar K_6^\pm = \displaystyle
-\frac{307\mp\sqrt{57751849}}{1997835840}\sin^4\bar h\cos\bar h
\big(4175357+6595320\cos(2\bar h)+4655076\cos(4\bar h)) \\
\hspace*{8.3cm}+2405704\cos(6\bar h)+1092063\cos(8\bar h)\big) \\[2mm]
\hspace*{2cm}\delta \bar V_6^\pm = \displaystyle
-\frac{1}{354816}\big(106772\cos\bar h+103950\cos(3\bar h)
+17(5775\cos(5\bar h)+5225\cos(7\bar h)\\[2mm]
\hspace*{4.7cm} +4389 \cos(9\bar h)+3059\cos(11\bar h))\big) \nn
\ea
\ee
The eigenvalues of the $(-)$--series are systematically smaller than those of the
$(+)$--series. Thus, the lowest eigenvalues are:
\footnote{For the eigenvalue $\lambda^+_1 = -8/13 = -0.615385$ the eigenfunction $\dk$
is singular at $h=0,\pi$ and $\delta\tilde K$ tends to a non zero constant at the endpoints,
so in this case the metric will be singular at both ends. }
$\lambda_1^-=-1.78,\,
\lambda_2^-=-0.50,\,
\lambda_3^-=1.83,\,
\lambda_4^-=5.03,\,
\lambda_5^-=9.07,\,
\lambda_2^+=13.07,\,
\lambda_6^-=13.98\ldots
$

Finally, let us ask what these results imply for the coupling of the Higgs field $h$
to the Goldstone bosons $\chi^\alpha$.
Following the notation of \cite{cgmpr} for the general parametrization
of Higgs couplings at low energies, we Taylor expand the function $K$ which appears in
\eqref{L_Simple} in powers of the shifted field $H=h-\langle h\rangle$
and the Goldstone boson Lagrangian reads:
\be
\label{fatmeh}
\frac{\upsilon^2}{2}
\left(1+2a\,\frac{H}{\upsilon}+b\,\frac{H^2}{\upsilon^2}+\cdots\right)
g_{\alpha\beta}\,\partial_\mu\chi^\alpha\partial^\mu\chi^\beta\ .
\ee
Here $\upsilon$=246 GeV is the weak scale and $\langle h\rangle$ is defined as the
position of the minimum of $V$.

The Gaussian fixed point is the basis of the perturbative treatment of linear scalar theory.
It describes $N$ free massless scalar fields. In particular, there is no Higgs VEV at the fixed point.
The eigenvalues of the linearized flow are just the canonical
dimensions of the operators appearing in the eigen--perturbation.
There are only two non-irrelevant perturbations: the mass and the quartic self--interaction.
A combination of these perturbations with suitable coefficients generates a Higgs VEV
and the coefficients $a$ and $b$ in \eqref{fatmeh} are both equal to one,
while all higher order couplings vanish.

At the cylindrical fixed point the Higgs and the Goldstone bosons are decoupled.
At one loop there are then non-irrelevant perturbations with $\delta\tilde K$ a polynomial
up to second order in $h$ and $\delta\tilde V$ a polynomial up to fourth order.
By suitable choice of coefficients one can generate a VEV for $h$.
At the same time, there are also perturbations describing quartic Higgs-Goldstone interactions
which have a similar form as in the standard model, but their strength is unrelated
to the VEV. Specifically, we parametrize the perturbations as
\bea
K(\rho) &=& k^{2}\left(\tilde K_*+\epsilon_i \,\delta \tilde{K}_i(\tilde h)\right) \\
V(\rho) &=& k^{4}\left(\tilde V_* +\epsilon'_i \,\delta\tilde V_i(\tilde h) \right)
\eea
where $\epsilon_i = c_i e^{t\lambda_i}$ and $\epsilon'_i = c'_i e^{t\lambda_i}$ with $c_i, c'_i$ being scale independent quantities and $i=1,2$. Here $\delta\tilde{K}_1(\tilde h), \delta\tilde{V}_1(\tilde h)$ are the two relevant and $\delta\tilde{K}_2(\tilde h), \delta\tilde{V}_2(\tilde h)$ the two marginal eigen--perturbations. Doing this the Higgs VEV will be given by
\vspace{-1mm}
\be
\langle \tilde h \rangle^2 = \frac{3(\epsilon'_1+2\epsilon'_2)}{64\pi^2\epsilon'_2} 
\vspace{-1mm}
\ee
and the couplings by
\be
a= -\frac{4\sqrt{6}\pi \sqrt{\epsilon'_1+2\epsilon'_2}\,\epsilon_2}{\sqrt{2\epsilon'_2(\epsilon_1-2\epsilon_2+\tilde{K}_*)-3\epsilon_2\epsilon'_1}} \ ;\qquad
b= -32\pi^2 \epsilon_2.
\ee
One can also compute the mass and the weak scale to get
\be
\frac{m^2}{k^2}\!= 128\pi^2 (\epsilon'_1+2\epsilon'_2)\ ;\qquad
\frac{v^2}{k^2} = \epsilon_1-\epsilon_2\left(2+\frac{3\epsilon'_1}{2\epsilon'_2}\right)+\tilde{K}_*.
\ee
Solving for $\epsilon_1$ and $\epsilon_2$ to set $a=b=1$, we find the following expression for the weak scale
\vspace{-1mm}
\be
\frac{v^2}{k^2} = \frac{3(\epsilon'_1+2\epsilon'_2)}{64\pi^2\epsilon'_2},
\vspace{-1mm}
\ee
so that the ratio of the mass squared to weak scale squared is 
\vspace{-1mm}
\be
\frac{m^2}{v^2} = \frac{8192\pi^4\epsilon'_2}{3}.
\vspace{-1mm}
\ee 
In the single-field approximation instead, we parametrize the perturbations as
\bea
K(\rho) &=& k^{2}\left(\tilde{K}_*+\epsilon_i \,\delta \tilde{K}_i(\tilde h)\right) \\
V(\rho) &=& k^{4}\left(\tilde V_* +\epsilon_i \,\delta\tilde V_i(\tilde h) \right)
\eea
with $\epsilon_i$ defined as before and $i=1, 2, 3, 4$. Here $(\delta \tilde{K}_i(\tilde h), \delta \tilde{V}_i(\tilde h))$ are the relevant or marginal eigen--perturbations in increasing order of eigenvalues. In terms of $\epsilon_i$ the Higgs VEV is given by
\be
\langle \tilde h \rangle^2 = \frac{3(32\epsilon_2-9\epsilon_3+64\epsilon_4)}{2048\pi^2\epsilon_4} \\
\ee
and the couplings are
\be
a = -\frac{6\sqrt{6}\pi \sqrt{32\epsilon_2-9\epsilon_3+64\epsilon_4}\,\epsilon_3}{81\epsilon_3^2+256\epsilon_4(\epsilon_1+\tilde{K}_*)-288\epsilon_2\epsilon_3-320\epsilon_3\epsilon_4} \ ;\qquad
b = -24\pi^2 \epsilon_3
\ee
The mass and the weak scale in this case will be
\be
\frac{m^2}{k^2}\! = 4\pi^2 (32\epsilon_2-9\epsilon_3+64\epsilon_4) \ ;\qquad
\frac{v^2}{k^2} = \epsilon_1+\frac{\epsilon_3(81\epsilon_3 - 320\epsilon_4-288\epsilon_2)}{256\epsilon_4}+\tilde{K}_*
\ee
As in the previous case we solve for $\epsilon_1$ and $\epsilon_3$ to set $a=b=1$. Doing this we find the following expressions for the mass and weak scale
\be
%\frac{m^2}{k^2}\! &=& \frac{3}{2}+ 128(\epsilon_2+2\epsilon_4)\pi^2 \\
\frac{m^2}{k^2}\! = \frac{3+ 256(\epsilon_2+2\epsilon_4)\pi^2}{2} \ ;\qquad
\frac{v^2}{k^2} = \frac{9+768(\epsilon_2+2\epsilon_4)\pi^2}{16384\pi^4\epsilon_4}
\ee
their ratio reads 
\be
\frac{m^2}{v^2} = \frac{8192\pi^4\epsilon_4}{3}.
\ee
We see that by adjusting the free parameters one can mimic the results of the Standard Model.

For the spherical fixed point in the \sfa\ there are two relevant eigen--perturbations one of which has $\delta\bar{K}\neq 0$. This will lead to modifications of Higgs couplings to Goldstone bosons. A linear combination of the two eigen--perturbations in the potential
$r\delta \bar V^-_1(\bar h)+\delta \bar V^-_2(\bar h)$, where $r=r_0\, \mathrm{exp}(t(\lambda^-_1-\lambda^-_2))$, will give rise to a non vanishing VEV, $\langle\bar h \rangle \equiv \theta$ given by
\be
\sin^2\!\theta = \frac{6-r}{7}
\ee
Using this notation we find the couplings to be
\bea
a &=& \cos\theta -\epsilon \sin^2\!\theta\,(3-4\sin^2\!\theta) \\
b &=& \cos(2\theta) -\epsilon \sin^2\!\theta \cos\theta\,(12-25\sin^2\!\theta)
\eea
where $\epsilon = \epsilon_0\, \mathrm{exp}(t\lambda^-_2)$ is some small parameter. The parameter $r$ can always be tuned to give a small enough $\theta$. On the other hand the mass and the weak scale will be given by
\bea
m^2 &=& -k^2 \tilde f_*^2 \,\frac{28}{3}\sin^2\!\theta\cos\theta\,\epsilon \\
v^2 &=& k^2 \tilde f_*^2 \left(\sin^2\!\theta-2\epsilon \sin^4\!\theta\cos\theta\right)
\eea
with the ratio being
\be
\frac{m^2}{v^2} = -\frac{28}{3}\cos\theta\,\epsilon
\ee
very roughly the values $r \approx 6$ and $\epsilon \approx - 0.028$
will reproduce the correct ratio $m^2/v^2 \approx 0.26$.
So also in this case we can get arbitrarily close to the result of the Standard Model.

\bigskip

\leftline{\bf Acknowledgements}
We thank A. Codello, T. Morris and O. Zanusso for discussions and A. Tonero for collaborating in the
initial stages of this work.

\goodbreak

\appendix

%%%%%%%%%%%%%%%%%%%%%%%%%%%%%%%%%%%%%%
\section{Calculation of beta functions}
%%%%%%%%%%%%%%%%%%%%%%%%%%%%%%%%%%%%%%

We give some details of the calculation of the two terms on the r.h.s. of \eqref{expans},
in the single--field approximation (for the adiabatic approximation one has to neglect the
$\zeta$'s in the r.h.s.).
The (infinite) matrices whose trace we must evaluate are
\be
({\cal P}^{-1}\dot{\cal R})^{i}_{j} =  \frac{\dot{R}_{k}+\zeta_{K} R_{k}+R'_{k}\dot{z}}{P+V'K'/2KJ}\,(\delta^i_j-\delta_{0}^{i} \delta^{0}_{j}) + \frac{\dot{R}_{k}+\zeta_J R_{k}+R'_{k}\dot{z}}{P+V''/J -V' J' /2J^2}\,\delta_{0}^{i} \delta^{0}_{j}
\ee
and
\be
(M{\cal P}^{-1}\dot{\cal R}{\cal P}^{-1})_{i}^{j} =  \frac{\dot{R}_{k}+ \zeta_{K} R_{k}+R'_{k}\dot{z}}{K (P+V'K'/2KJ)^2}\,M_{im}g^{mj} + \frac{\dot{R}_{k}+\zeta_J R_{k}+R'_{k}\dot{z}}{J(P+V''/J -V' J' /2J^2)^2}\,M_{im}\,\delta_{0}^{m} \delta_{0}^{j}
\ee
These can be evaluated using the following general formulas for the trace of a function
of the Laplacian:
\bea
\mathrm{Tr}[W(\Delta)] &=& \sum_{\lambda} W(\lambda)=
\frac{1}{(4\pi)^{\frac{d}{2}}} \sum_{n= 0}^{\infty} B_{2n}(\Delta)\, Q_{\frac{d}{2}-n}(W)
\eea
\bea
\mathrm{Tr}[W(\Delta)\,\partial_t \Delta] &=&
-\frac{1}{(4\pi)^{\frac{d}{2}}} \sum_{n= 0}^{\infty} \partial_t B_{2n}(\Delta)\, Q_{\frac{d}{2}-n+1}(W)
\eea
where $B_{2n}$ are the coefficients appearing in the heat kernel expansion
\be
\mathrm{Tr}\, e^{-s\Delta} = \frac{1}{(4\pi)^{\frac{d}{2}}} \sum_{n= 0}^{\infty} B_{2n}(\Delta) s^{-\frac{d}{2}+n}
\ee
and the $Q$-functionals are given (for $m$ a non-negative integer) by
$Q_m(W)=\frac{1}{\Gamma(m)}\int_{0}^{\infty}\!\! dz z^{m-1} W(z)$.
For convenience we choose the optimized cutoff $R_{k}(z)=(k^2-z)\theta(k^2-z)$
\cite{optimized}, for which one can easily calculate
\bea
Q_n \left(\frac{\dot{R}_{k}+\eta R_{k}}{(P_k +q)^l}\right) &=&
\frac{k^{2(n-l+1)}}{\Gamma(n+1)}\frac{2+\frac{\eta}{n+1}}{(1 +\tilde{q})^l}
\eea

With the aid of these formulas one gets
\bea
\label{TrPR}
\mathrm{Tr}_0[{\cal P}^{-1}\dot{\cal R}] %&=&
&=& 2c_d k^{d} \int \!\! d^{d}x \left[\left(1+\frac{\zeta_K}{d+2}\right)\frac{N-1}{1+\frac{V'K'}{2KJk^2}} +\left(1+\frac{\zeta_J}{d+2}\right)\frac{1}{1+\frac{V''}{Jk^2}-\frac{V'J'}{2Jk^2}}\right]
\eea
and
{\setlength\arraycolsep{2pt}
\bea
\label{TrPMPR}
\mathrm{Tr}_0[{\cal P}^{-1}M{\cal P}^{-1}\dot{\cal R}] %&=&
&=&  2c_d k^{d-2} \!\int \!\! d^{d}x \left[\left(1+\frac{\zeta_K}{d+2}\right)\frac{N-1}{4}
\frac{KK'J'+K^{\prime 2}J-2KK''J}{K^2J\left(1+\frac{V'K'}{2KJk^2}\right)^2} \right]
\partial^\mu \varphi^0 \partial_\mu \varphi^0
\nn \\
&+&  2c_d k^{d-2} \!\int \!\! d^{d}x \left[\left(1+\frac{\zeta_K}{d+2}\right)\frac{N-2}{4}
\frac{4KJ-K^{\prime 2}}{KJ\left(1+\frac{V'K'}{2KJk^2}\right)^2}\right]
\partial^\mu\varphi^\alpha\partial_\mu\varphi^\beta g_{\alpha\beta}
\nn \\
&+& 2c_d k^{d-2} \!\int \!\! d^{d}x \left[\left(1+\frac{\zeta_J}{d+2}\right)
\frac{KK'J'+K^{\prime 2}J-2KK''J}{4KJ^2\left(1+\frac{V''}{Jk^2}-\frac{V'J'}{2Jk^2}\right)^2}\right]
\partial^\mu \varphi^\alpha \partial_\mu \varphi^\beta g_{\alpha\beta}
\eea}%
where $c_d= \frac{1}{(4\pi)^{d/2}\Gamma(d/2+1)}$ and by $\mathrm{Tr}_0$ we mean the $B_0$ term in the heat kernel expansion.
In the final equation we have used
\be
J^{-1}M_{00} = \frac{KK'J'+K^{\prime 2}J-2KK''J}{4KJ^2} \,
\partial^\mu \varphi^\alpha \partial_\mu \varphi^\beta g_{\alpha\beta}
\ee
and
\be
K^{-1}\,M_{\alpha\beta}\,g^{\alpha\beta} = \frac{N-1}{4}\frac{KK'J'+ K^{\prime 2}J-2KK'' J}{K^2 J} \,
\partial^\mu \varphi^0 \partial_\mu \varphi^0 +\frac{N-2}{4}\frac{4KJ-K^{\prime 2}}{KJ} \,\partial^\mu \varphi^\alpha \partial_\mu \varphi^\beta g_{\alpha\beta}
\ee

Collecting, one obtains
{\setlength\arraycolsep{2pt}
\bea
\label{BetaJ}
\zeta_J \,&=&  \frac{2c_d k^{d-2}}{J}
\left(1+\frac{\zeta_K}{d+2}\right)\frac{N-1}{4}
\frac{K K'J'+ K^{\prime 2}J-2KK'' J}{K^2 J(1+V'K'/2KJk^2)^2}
\\
\label{BetaK}
\zeta_K &=&  \frac{2c_d k^{d-2}}{K} \left[\left(1+\frac{\zeta_K}{d+2}\right)
\frac{N-2}{4}
\frac{4K J-K^{\prime 2}}{K J(1+V'K'/2KJk^2)^2}
\right. \nn
\\
&&
\left. \hspace{12mm} +\left(1+\frac{\zeta_J}{d+2}\right)
\frac{K K' J'+K^{\prime 2}J-2 K K'' J}{4 K J^2 (1+V''/Jk^2 -V'J'/2J^2k^2)^2}\right]
\\
\label{BetaV}
\zeta_V &=&  \frac{c_d k^{d}}{V}
\left[\left(1+\frac{\zeta_K}{d+2}\right)\frac{N-1}{1+V'K'/2KJk^2} +\left(1+\frac{\zeta_J}{d+2}\right)\frac{1}{1\! +\! V''/Jk^2 \! -\! V'J'/2J^2k^2}\right]
\eea}%
This can be rewritten in terms of dimensionless variables as in section II.

\end{document}